\documentclass[aps,prl,twocolumn,showpacs,superscriptaddress,floatfix]{revtex4}
\usepackage{bm}
\usepackage{color}
\usepackage{amssymb}
\usepackage{graphicx}
\usepackage{amsmath}
\usepackage{dcolumn}
\usepackage{bm}
\usepackage[german,english]{babel}

\begin{document}

\title{Effect of Point Defects on the Optical and Transport Properties of MoS$_2$ and WS$_2$}

\author{Shengjun Yuan}
\email{s.yuan@science.ru.nl}
\affiliation{Radboud University of Nijmegen, Institute for Molecules and Materials,
Heijendaalseweg 135, 6525 AJ Nijmegen, The Netherlands}
\author{Rafael Rold\'an}
\email{rroldan@icmm.csic.es}
\affiliation{Instituto de Ciencia de Materiales de Madrid, CSIC, Cantoblanco E28049
Madrid, Spain}
\author{M. I. Katsnelson}
\affiliation{Radboud University of Nijmegen, Institute for Molecules and Materials,
Heijendaalseweg 135, 6525 AJ Nijmegen, The Netherlands}
\author{Francisco Guinea}
\affiliation{Instituto de Ciencia de Materiales de Madrid, CSIC, Cantoblanco E28049
Madrid, Spain}
\pacs{}
\date{\today }

\begin{abstract}
Imperfections in the crystal structure, such as point defects, can strongly modify the optical and transport properties of materials. Here, we study the effect of point defects on the optical and DC conductivities of single layers of semiconducting transition metal dichalcogenides with the form $M$S$_2$, where $M$=Mo or W. The electronic structure is considered within a six bands tight-binding model, which accounts for the relevant combination of $d$ orbitals of the metal $M$ and $p$ orbitals of the chalcogen $S$. We use the Kubo formula for the calculation of the conductivity in samples with different distributions of disorder. We find that $M$ and/or S defects create mid-gap states that localize charge carriers around the defects and which modify the optical and transport properties of the material, in agreement with recent experiments. Furthermore, our results indicate a much higher mobility for $p$-doped WS$_2$ in comparison to MoS$_2$.
\end{abstract}

\maketitle

{\it Introduction---} The mobility of current single layer crystals of transition metal dichalcogenides (TMD) is highly dependent on the screening environment and is limited by the presence of defects in the samples. The existence of defects in the chemical and structural composition of those materials can influence their optical and transport properties, as revealed by recent experimental results.  A broad peak at $\sim700$ nm ($\sim 1.77$ eV) in the optical spectrum of bilayer MoS$_2$ has been associated to impurities \cite{WX13} whereas the mobility of multilayer samples has been shown to highly depend on the substrate and dielectric effects \cite{BF13}. Vacancy defects in the crystal, which can be created by means of thermal annealing and $\alpha$-particle \cite{TW13} or electron beam irradiation \cite{ZI13,QiuH2013}, trap free charge carriers and localize excitons, leading to new peaks in the photoluminescence spectra \cite{TW13}. Recent experiments \cite{QiuH2013} show that the density of sulphur vacancies in MoS$_2$ is of the order of $10^{13}$ cm$^{-2}$, corresponding to an average defect distance about 1.7 nm. The existence of line defects, which separate patches or islands where the layer direction is opposite to its surrounding, can lead to changes in the carrier mobility \cite{ES13}, and the importance of short-range disorder has been proposed as the main limitation for the mobility of chemical vapor deposition (CVD) grown single-layer MoS$_2$ \cite{ZA14,SE14}.

Therefore, it is necessary to understand the effect of impurities in the optical and transport properties of TMD, as a first step to exploit the controlled creation of defects as a route to manipulate their electronic properties. There are several theoretical works which have studied this problem using {\it ab initio} methods \cite{AC11,MH11,WP12,KK12,GH13,ZZ13,ES13,LR13}. However, the simulation of realistic disordered samples of TMD with a random distribution of defects are extremely expensive computationally for density functional theory (DFT) methods, since they require a very large unit cell in the calculation. In this paper we follow an alternative route and perform a systematic study of the density of states, optical and DC conductivities of single-layers of MoS$_2$ and WS$_2$ in the presence of point defects, by means of a {\it real space} tight-binding (TB) model for large systems, containing millions of atoms. In our simulations, Mo/W and S point defects are introduced by elimination of atoms which are randomly distributed over the sample. This method allows us to study point defects such as unreconstruced vacancies, chemically bonded atoms or molecules, and strong substitutional defects which prevent the electronic hopping to the neighbors. We also consider clusters of point defects. We use a TB model that considers the relevant orbital contribution in the valence and conduction bands, as well as the effect of spin-orbit coupling (SOC) \cite{CG13,RG14}. The optical and electronic properties are obtained numerically by using the tight-binding propagation methods (TBPM) \cite{YRK10,WK10,YRRK11,Yuan2012}. Our results show that point defects create midgap states whose energy depends on the specific impurities. We show that optical transitions involving the impurity bands lead to a background contribution in the  photoconductivity at low energies, in agreement with recent experiments \cite{MH10}. We further calculate the DC conductivity of disordered MoS$_2$ and WS$_2$, finding that the impurity states do not contribute to the conductivity within the gap, whereas they lead to a depletion of it outside the gap.

\begin{figure*}[t]
\begin{center}
\mbox{
\includegraphics[width=5.6cm]{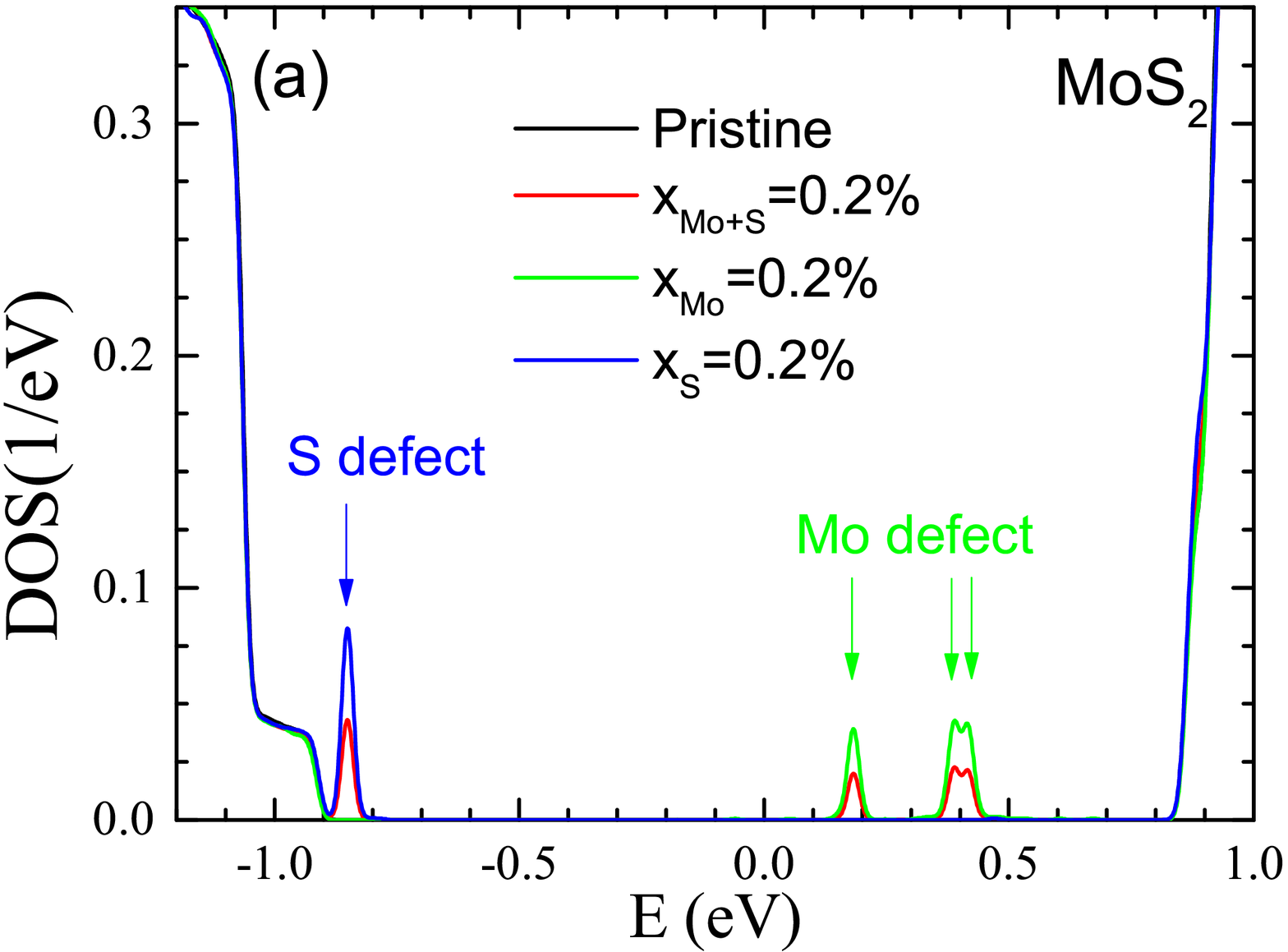}
\includegraphics[width=5.6cm]{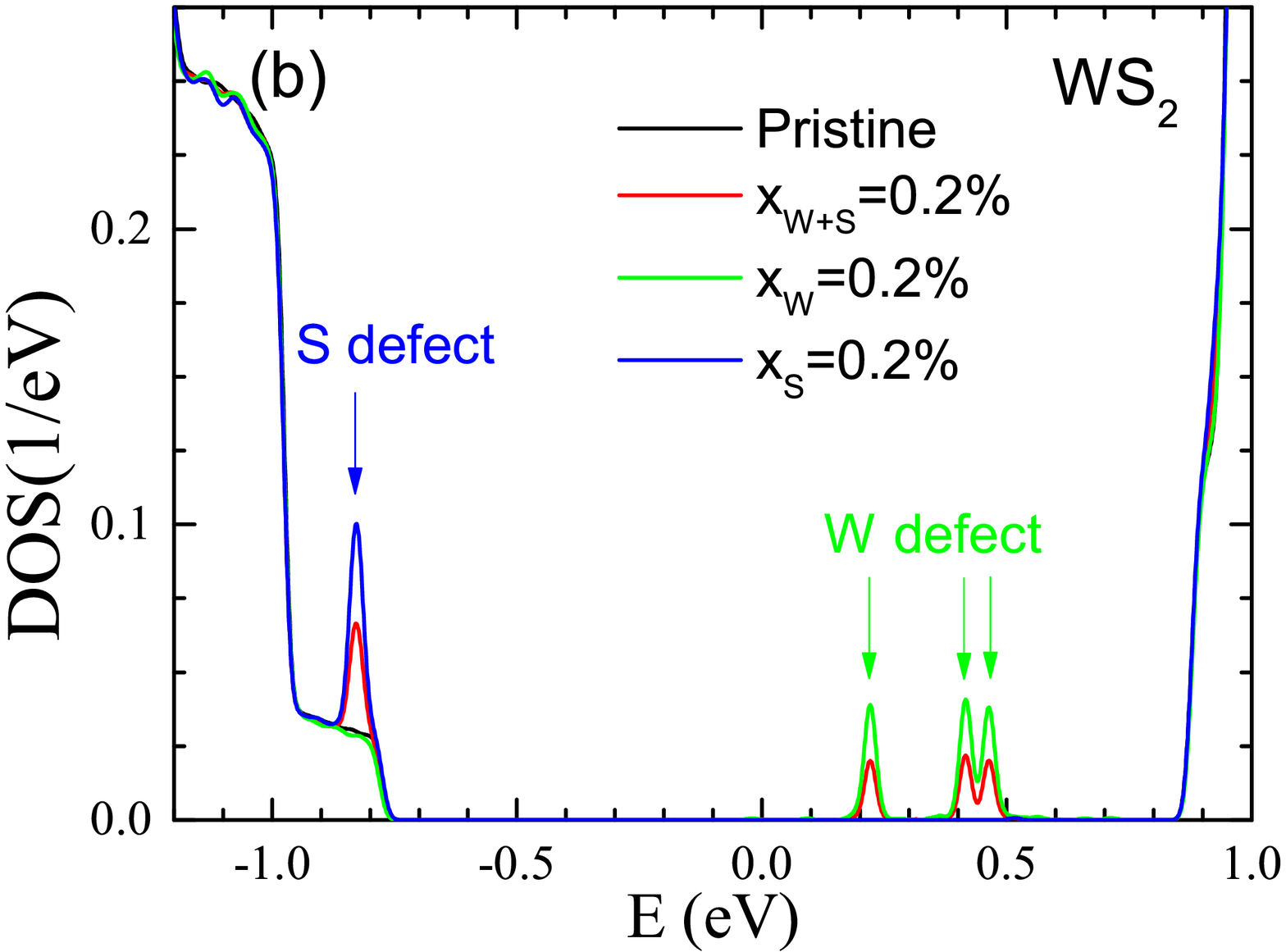}
\includegraphics[width=5.6cm]{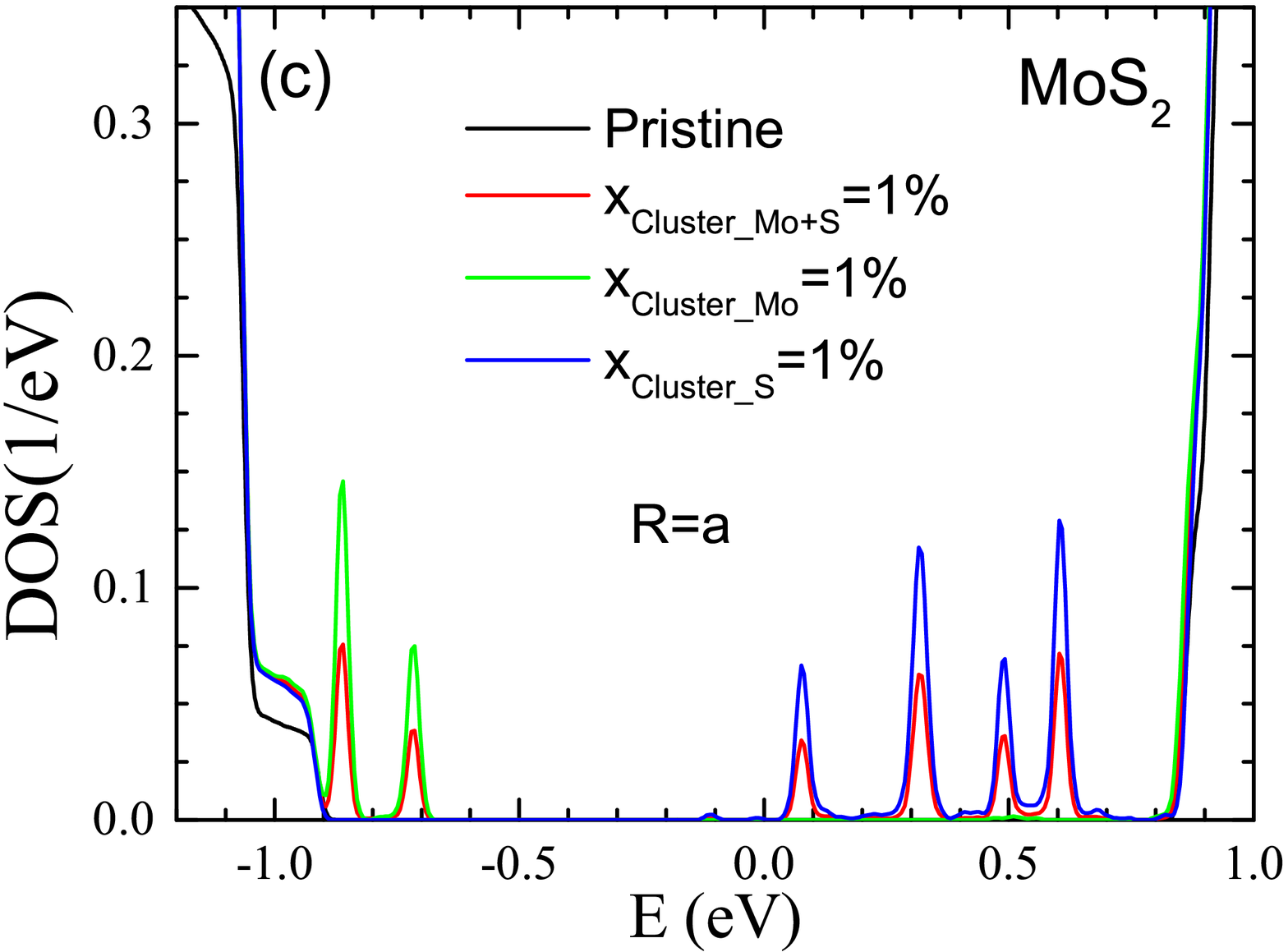}
}
\mbox{
\includegraphics[width=5.6cm]{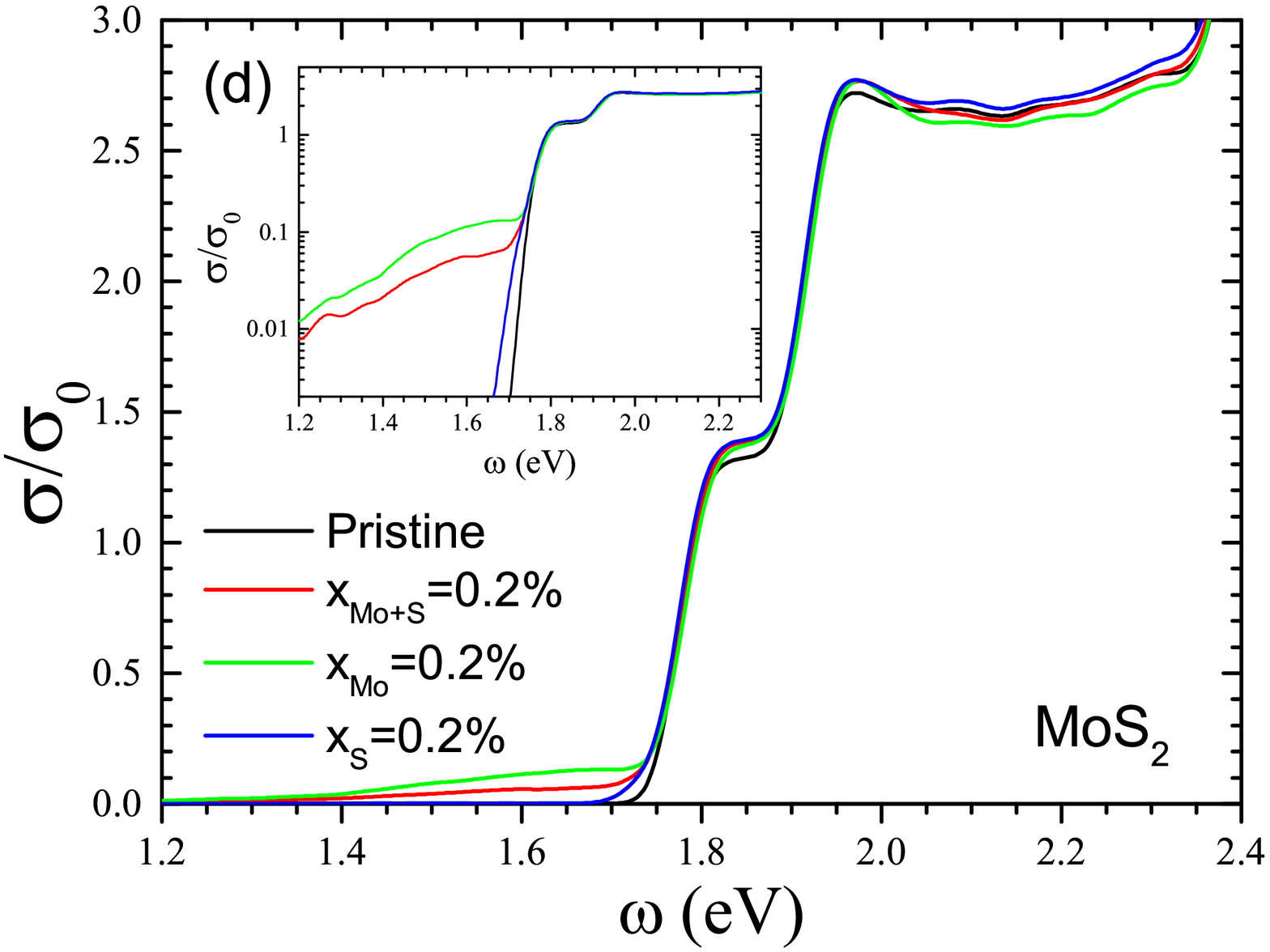}
\includegraphics[width=5.6cm]{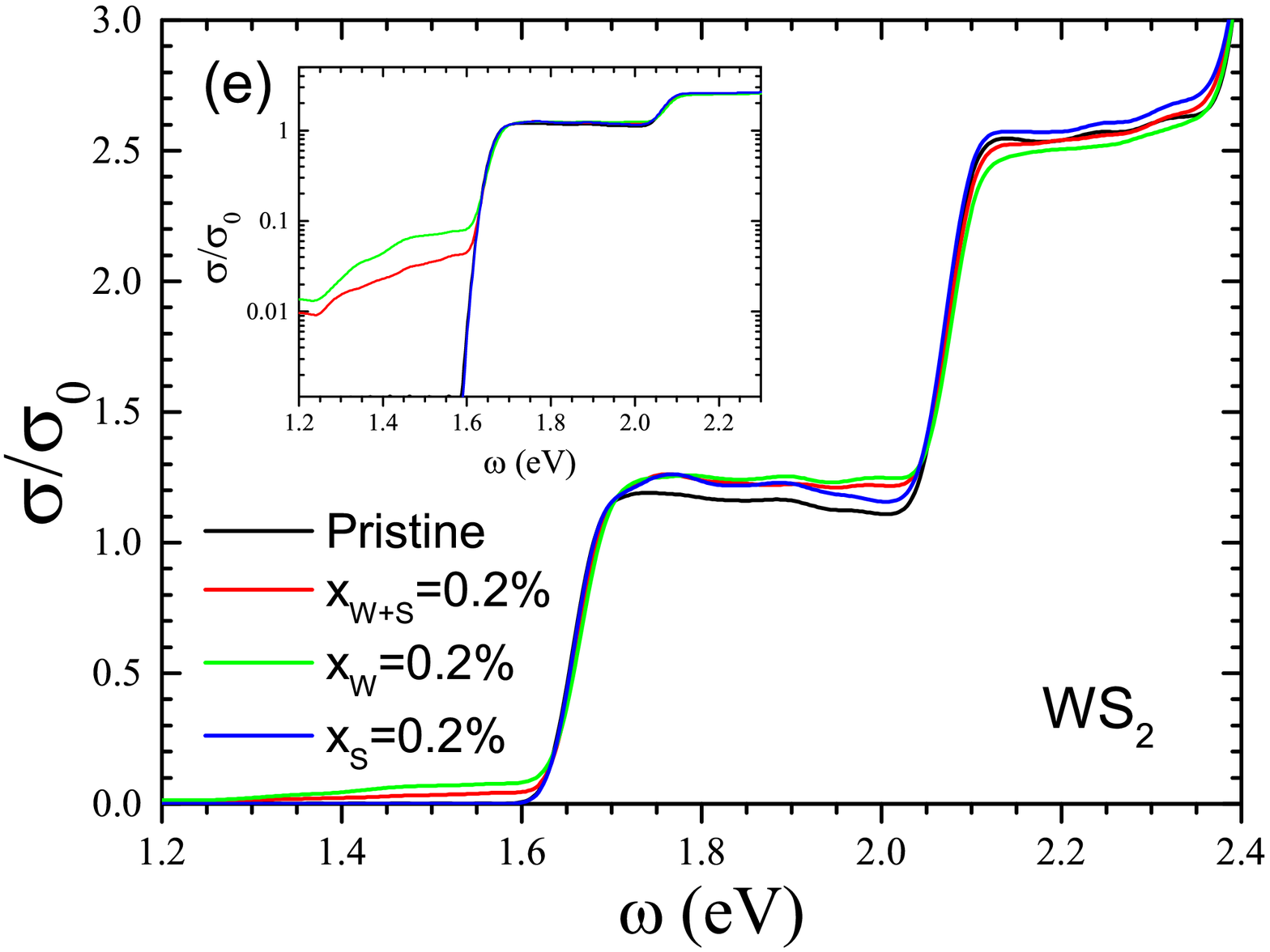}
\includegraphics[width=5.6cm]{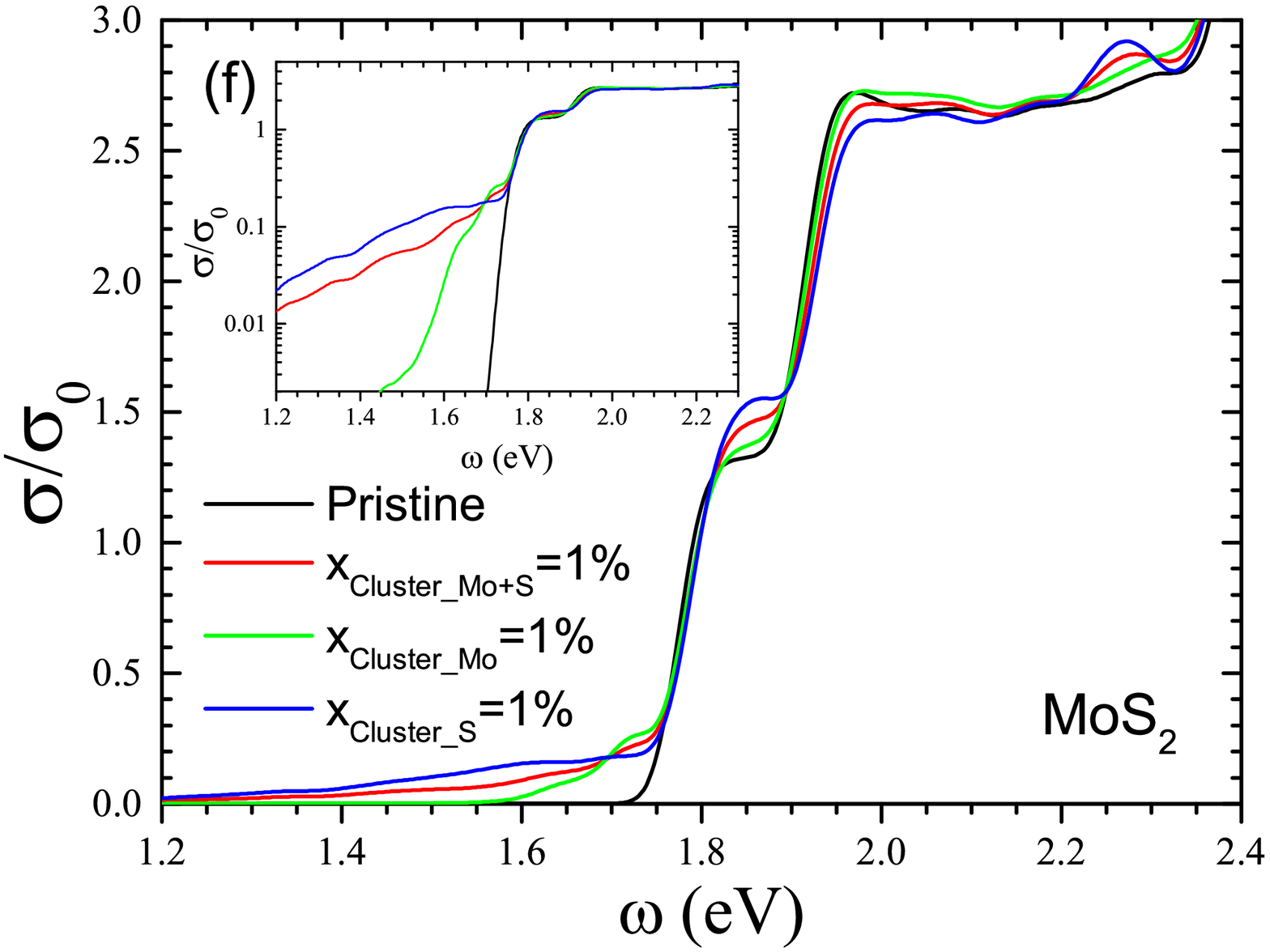}
}
\end{center}
\caption{DOS of MoS$_2$ (a) and WS$_2$ (b) for different types of single point defects, and for MoS$_2$ (c) with clusters of defects with $R=a$ (the labels in the subscript of the concentration x indicate the center of the cluster). The peaks in the DOS associated to midgap bands due to S or Mo/W defects are marked by arrows. Optical conductivity (in units of $\sigma _{0}=\pi e^{2}/2h$) for MoS$_2$ (d) and WS$_2$ (e) with single defects, and for MoS$_2$ with clusters of defects, for the same concentration of defects as in (a)-(c). The insets show the same plots in a logarithmic scale. }
\label{Fig:DOS&Optical}
\end{figure*}

\begin{figure*}[t]
\begin{center}
\mbox{
\includegraphics[width=5.6cm]{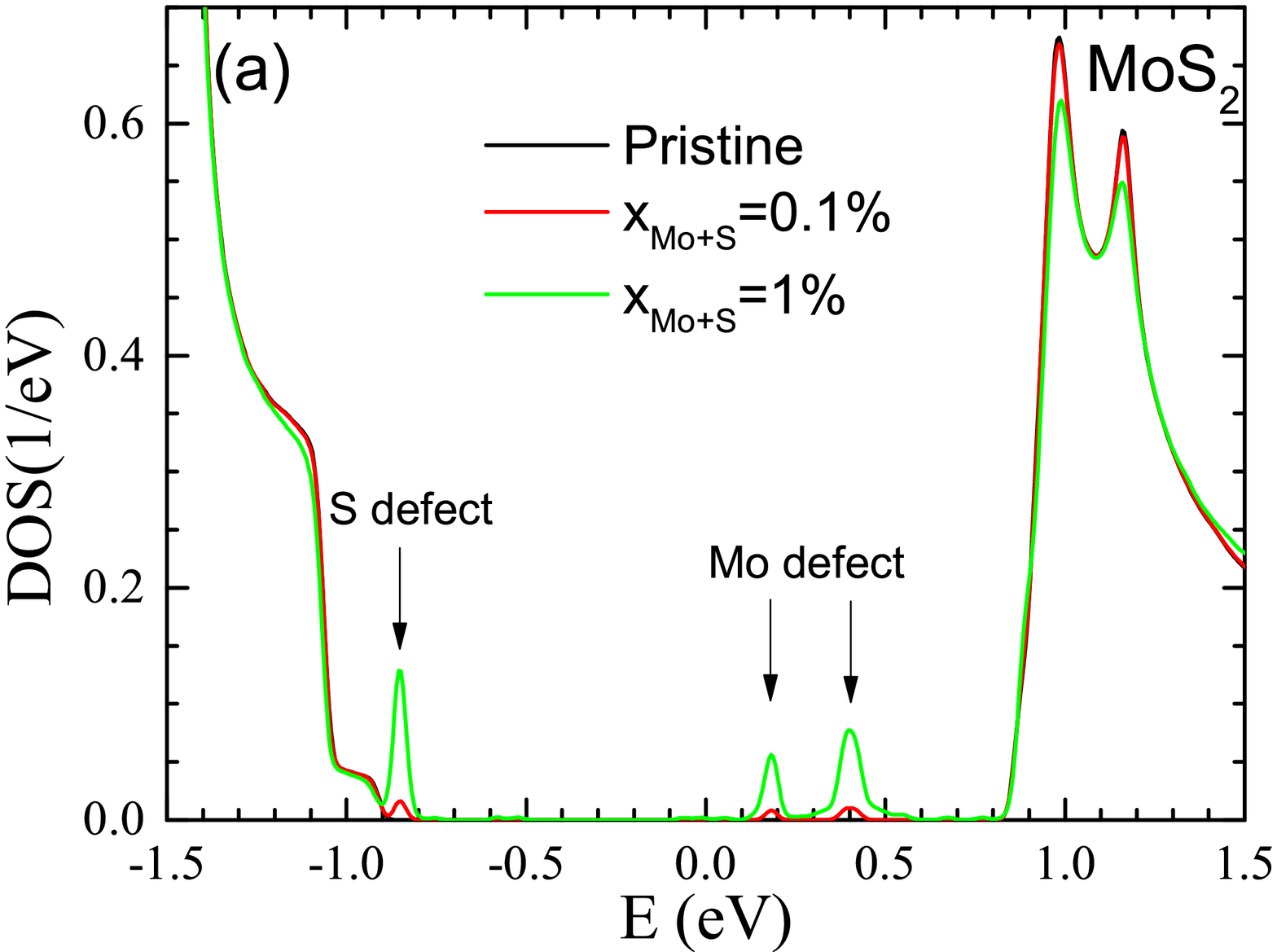}
\includegraphics[width=5.6cm]{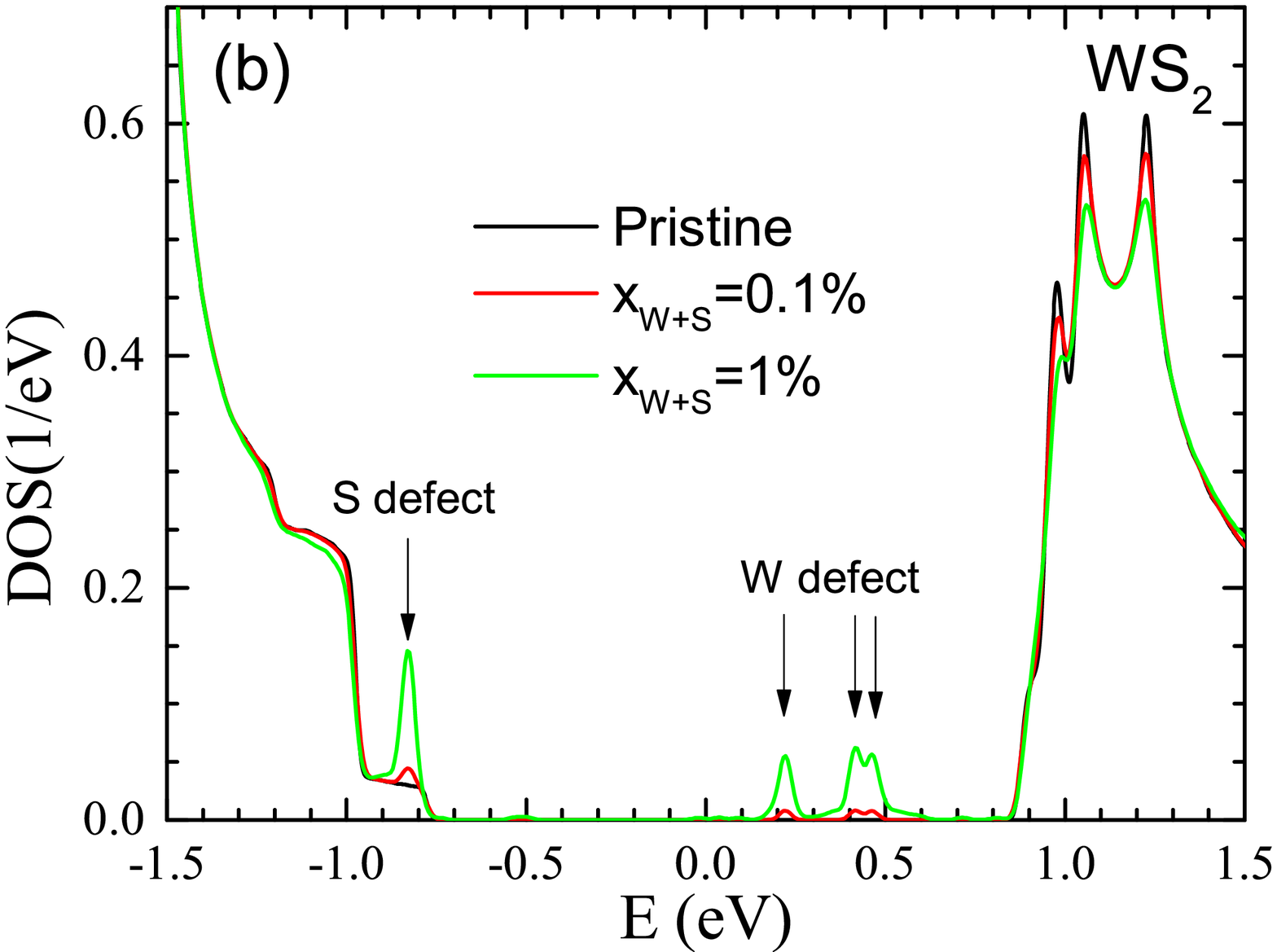}
\includegraphics[width=5.6cm]{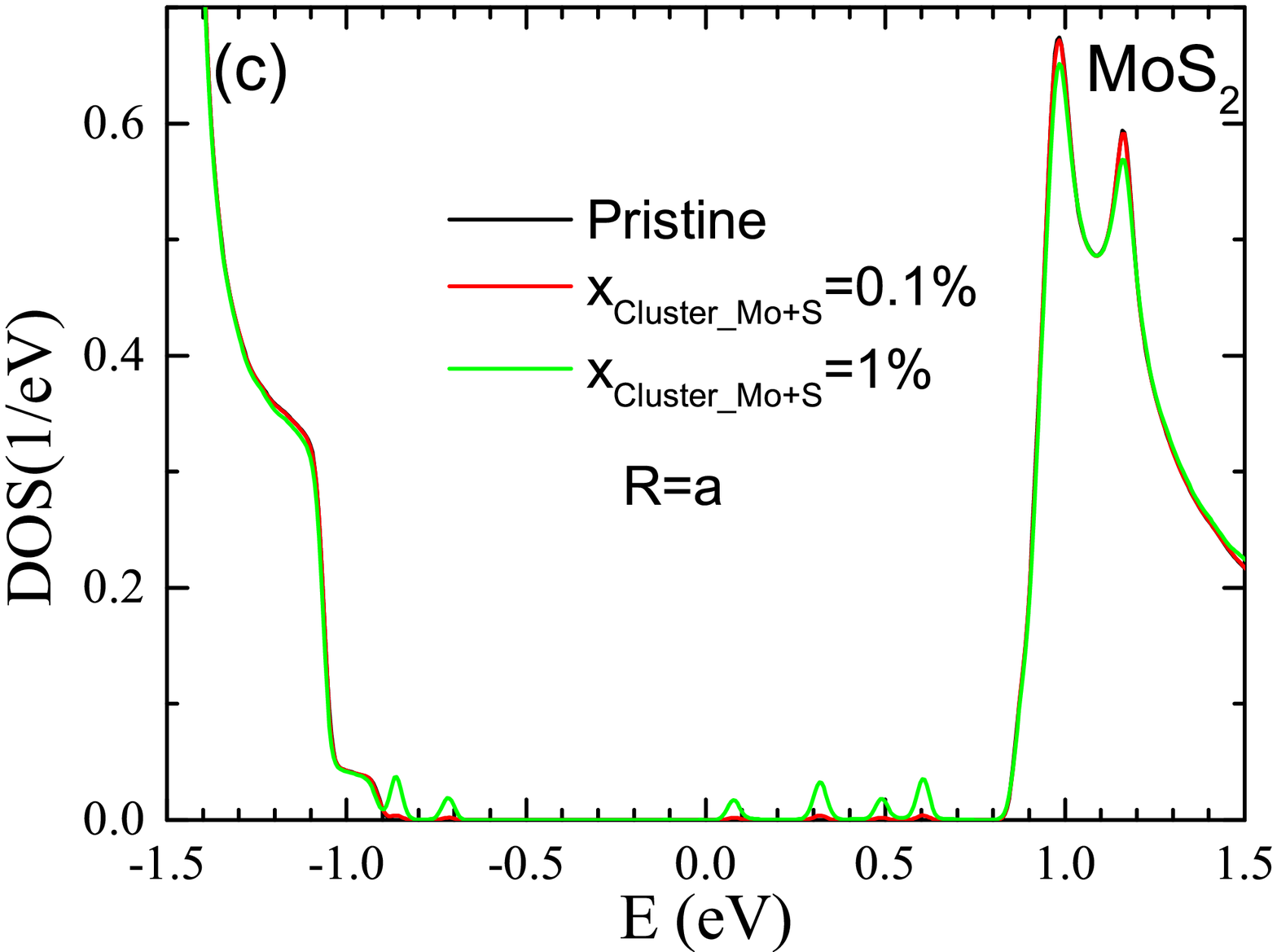}
}
\mbox{
\includegraphics[width=5.6cm]{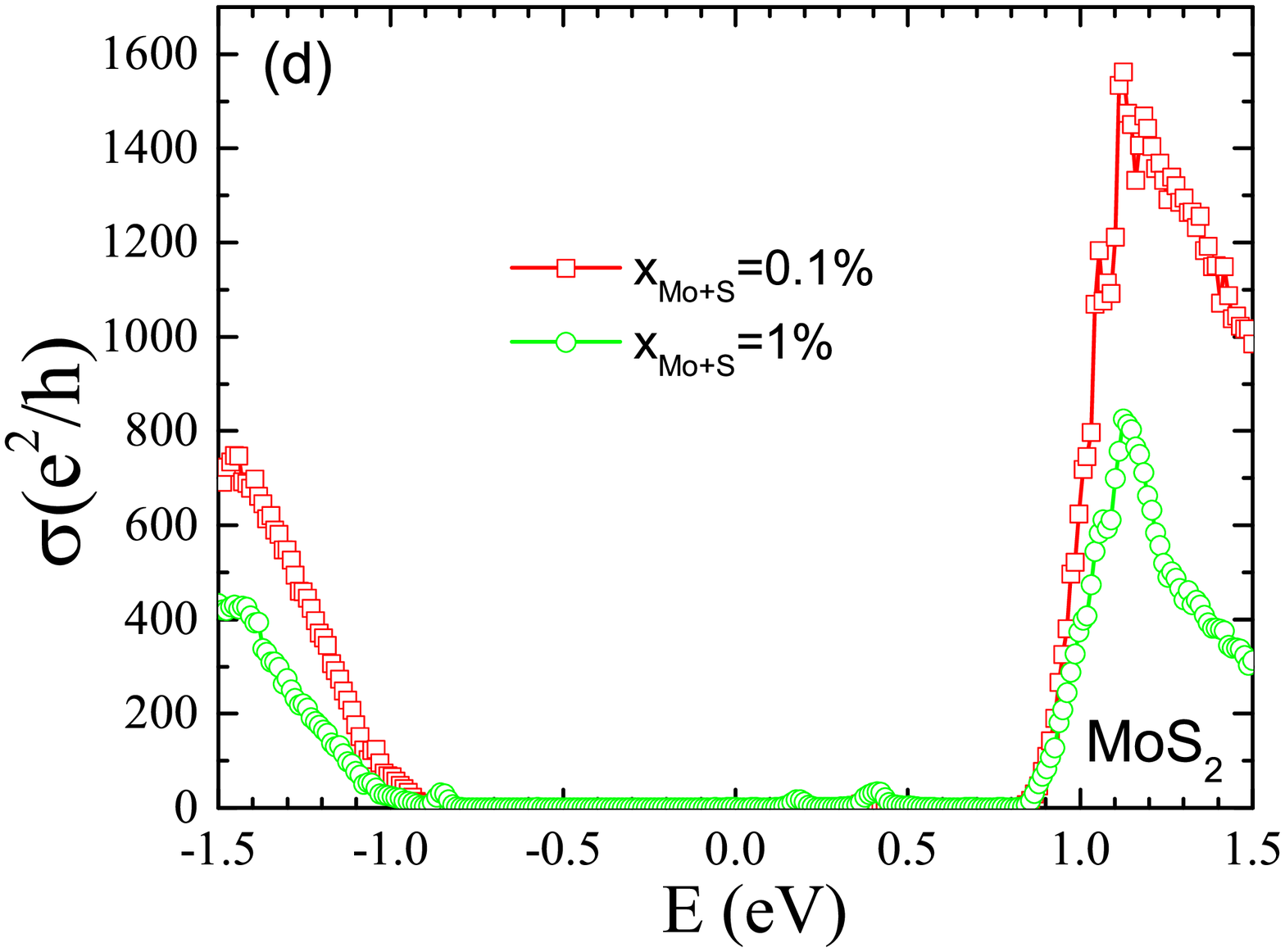}
\includegraphics[width=5.6cm]{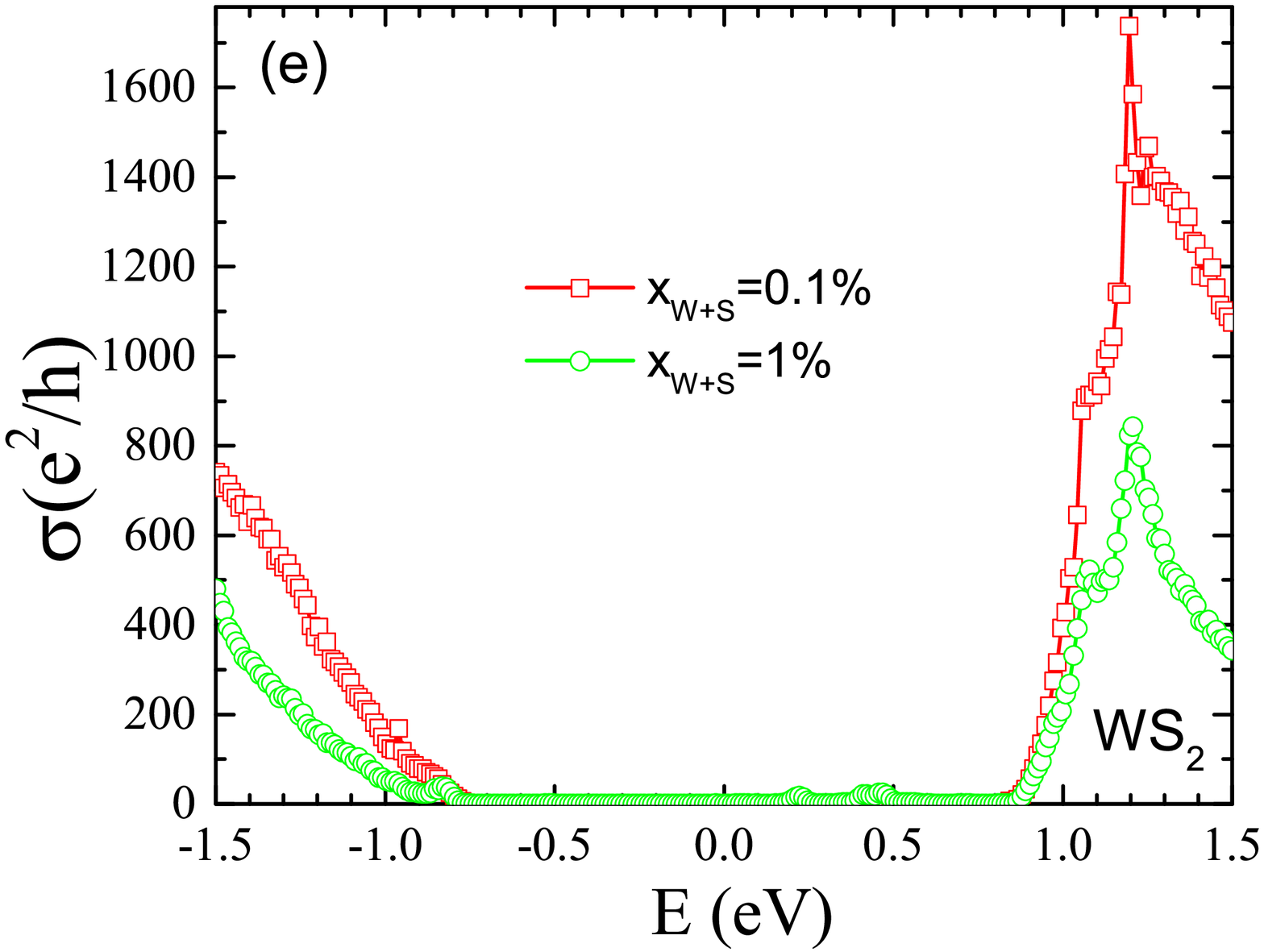}
\includegraphics[width=5.6cm]{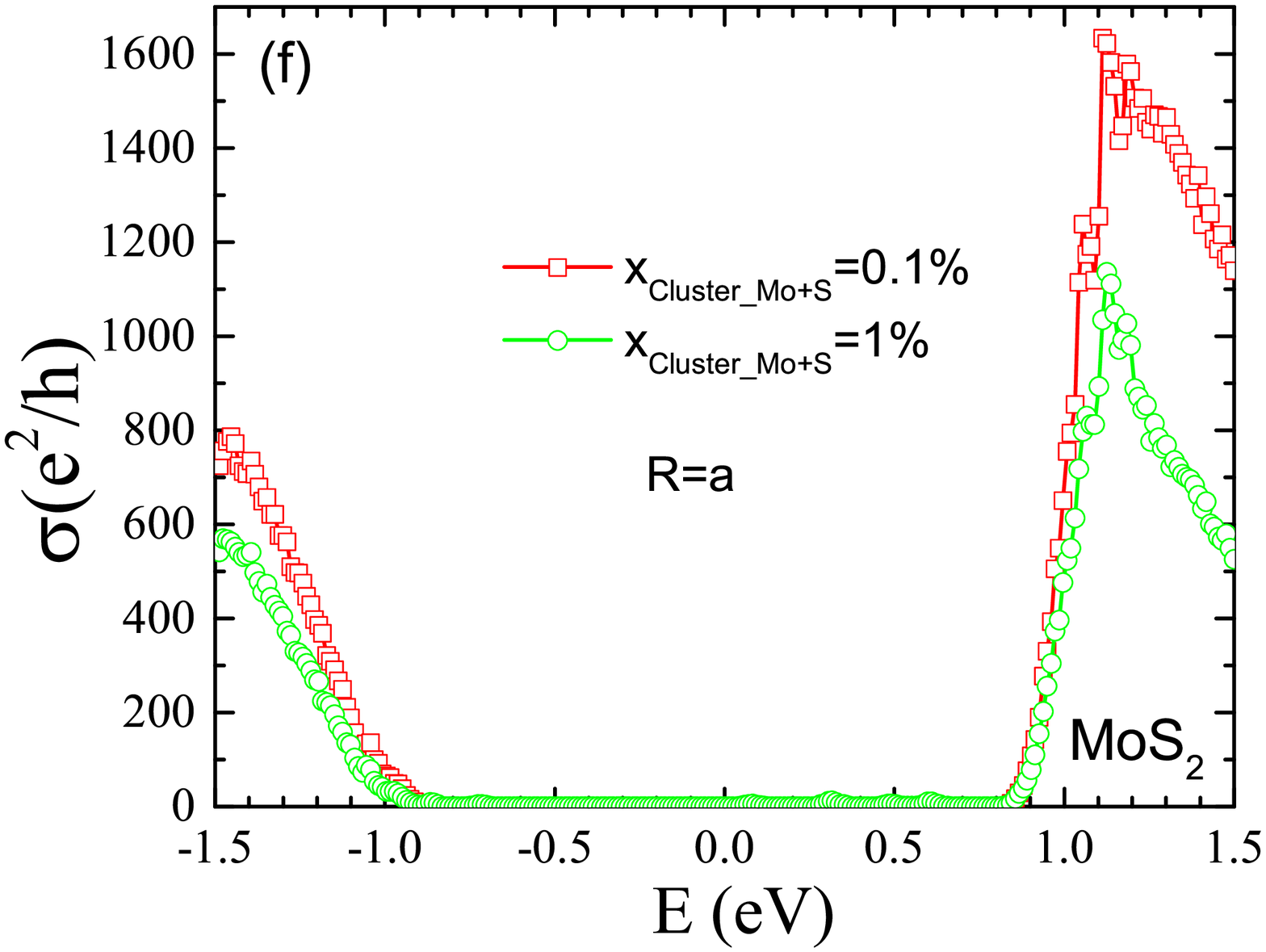}
}
\end{center}
\caption{Top panels: DOS of (a) MoS$_2$ and (b) WS$_2$ with single defects, and of (c) MoS$_2$ with clusters of defects with radius $R=a$. Panels (d)-(f) show the DC conductivity as a function of doping energy for the same concentrations of defects as in (a)-(c). 
}
\label{Fig:DC}
\end{figure*}

\begin{figure*}[t]
\begin{center}
\mbox{
\includegraphics[width=5.6cm]{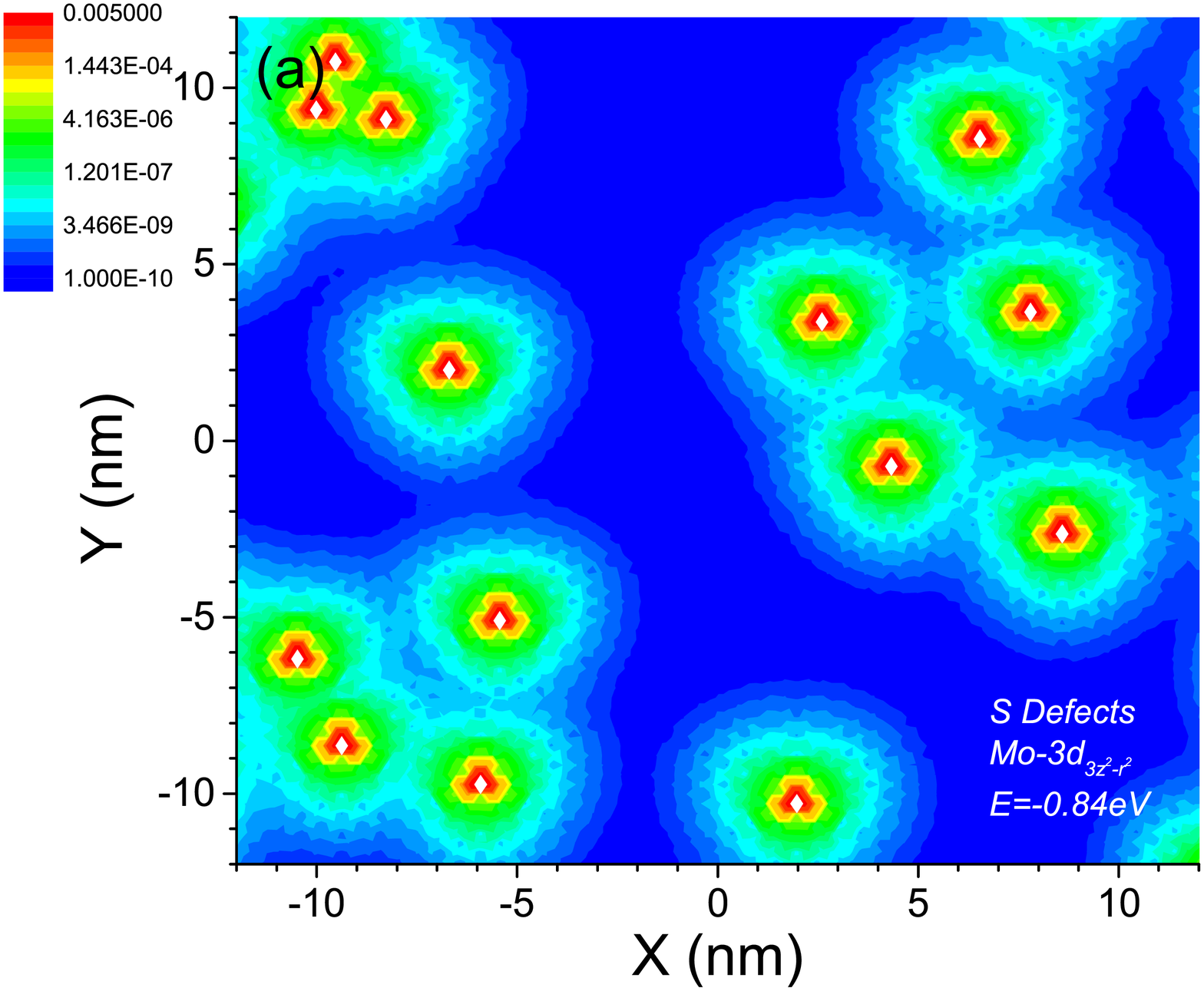}
\includegraphics[width=5.6cm]{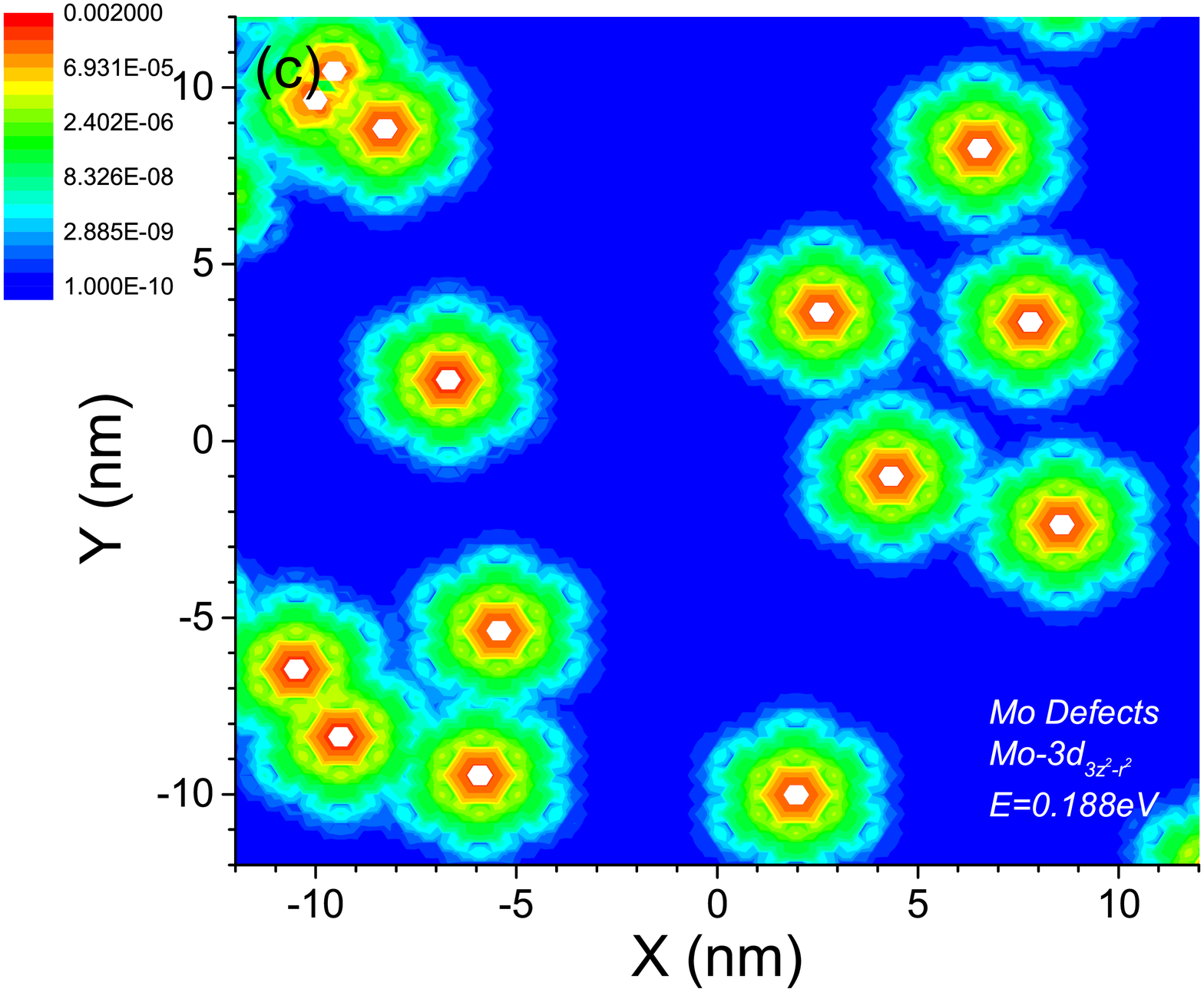}
\includegraphics[width=5.6cm]{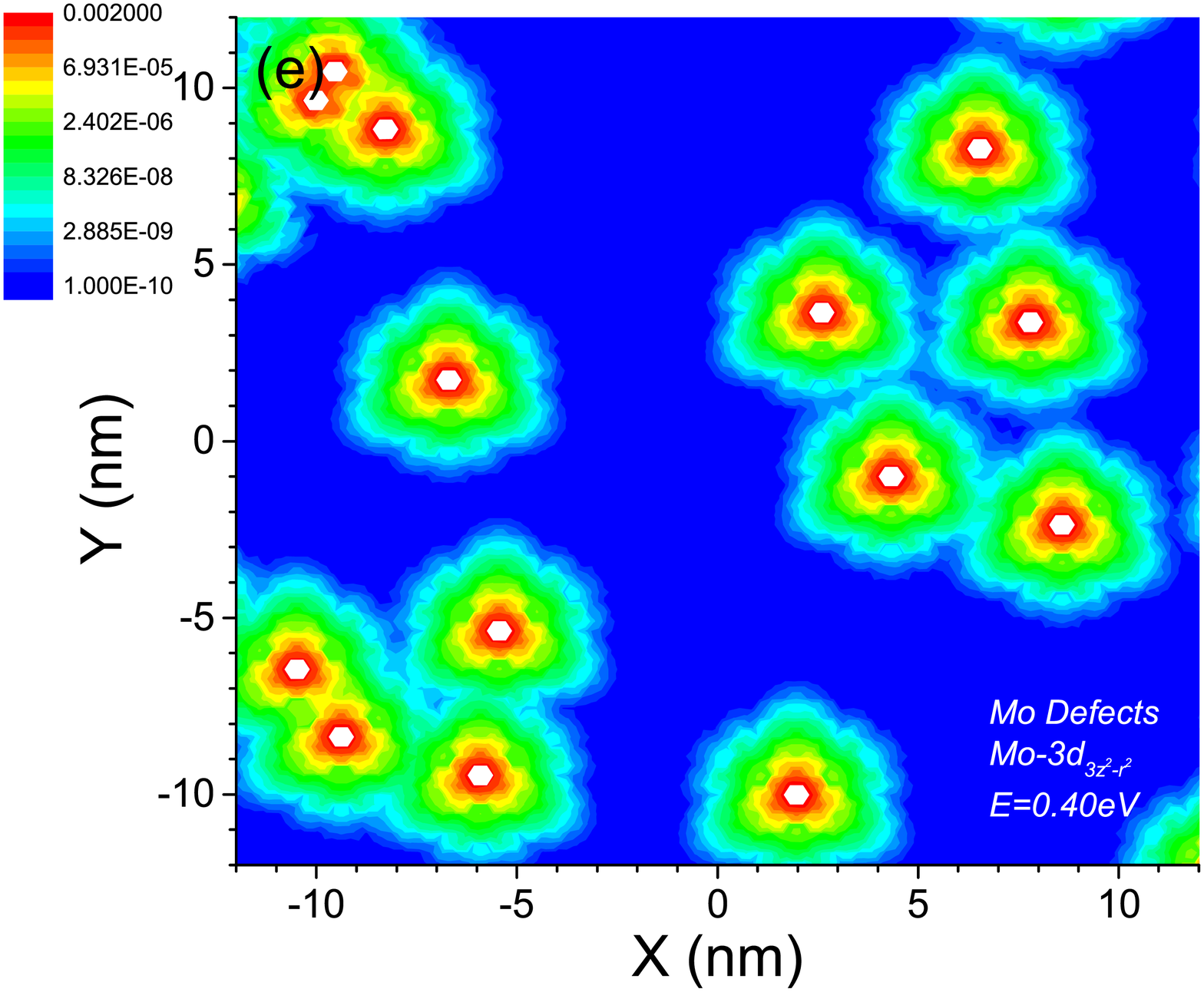}

}
\mbox{
\includegraphics[width=5.6cm]{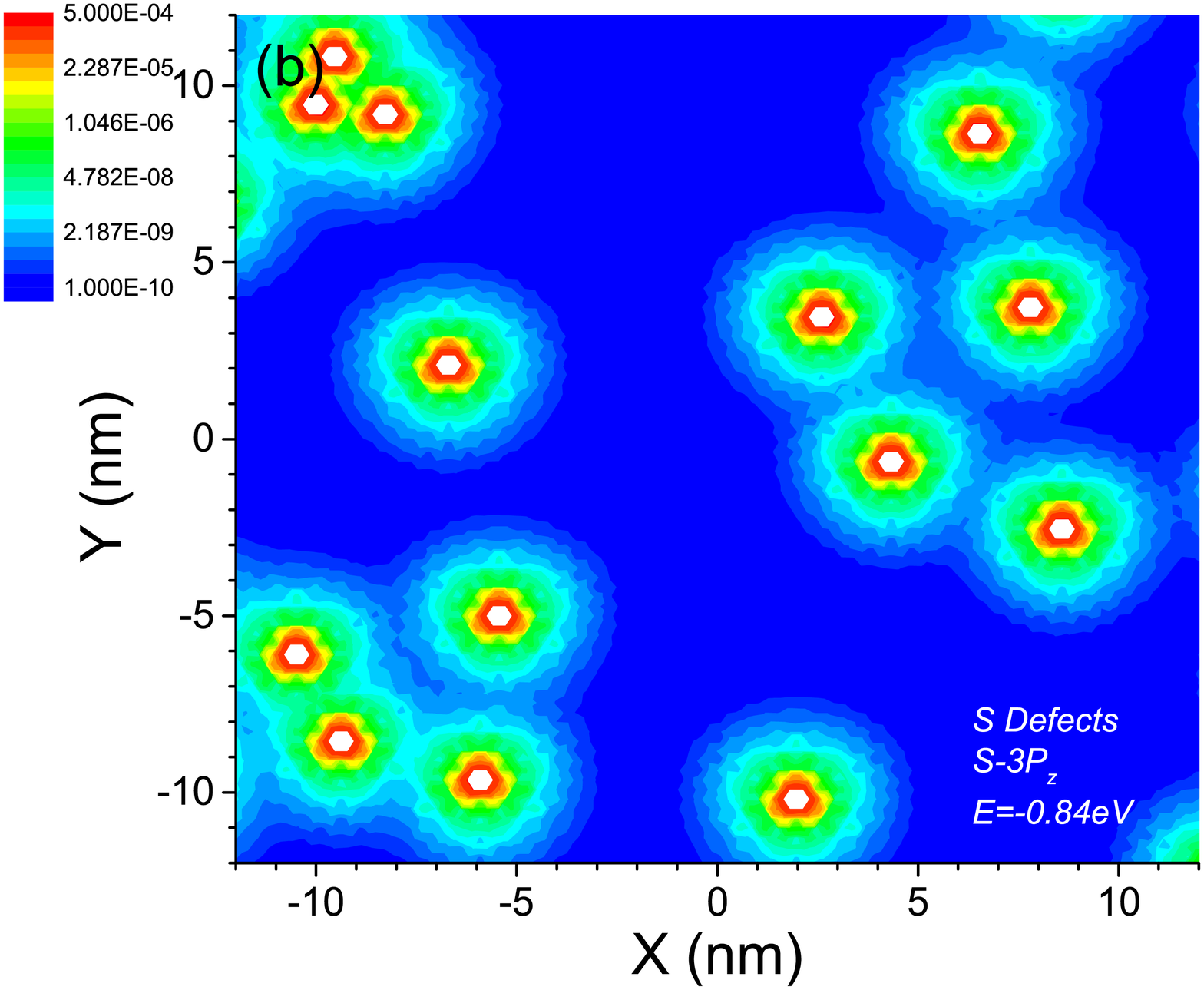}
\includegraphics[width=5.6cm]{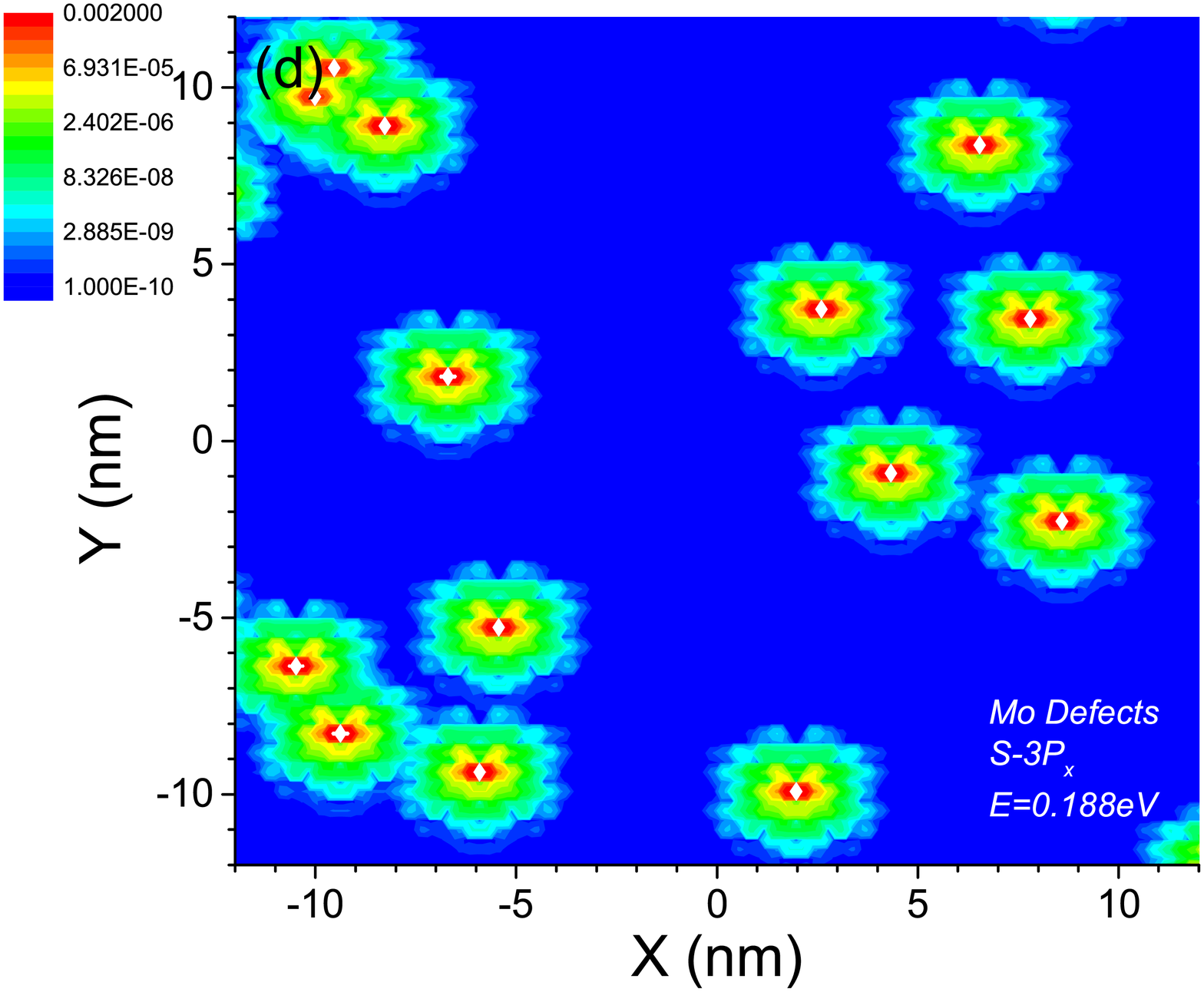}
\includegraphics[width=5.6cm]{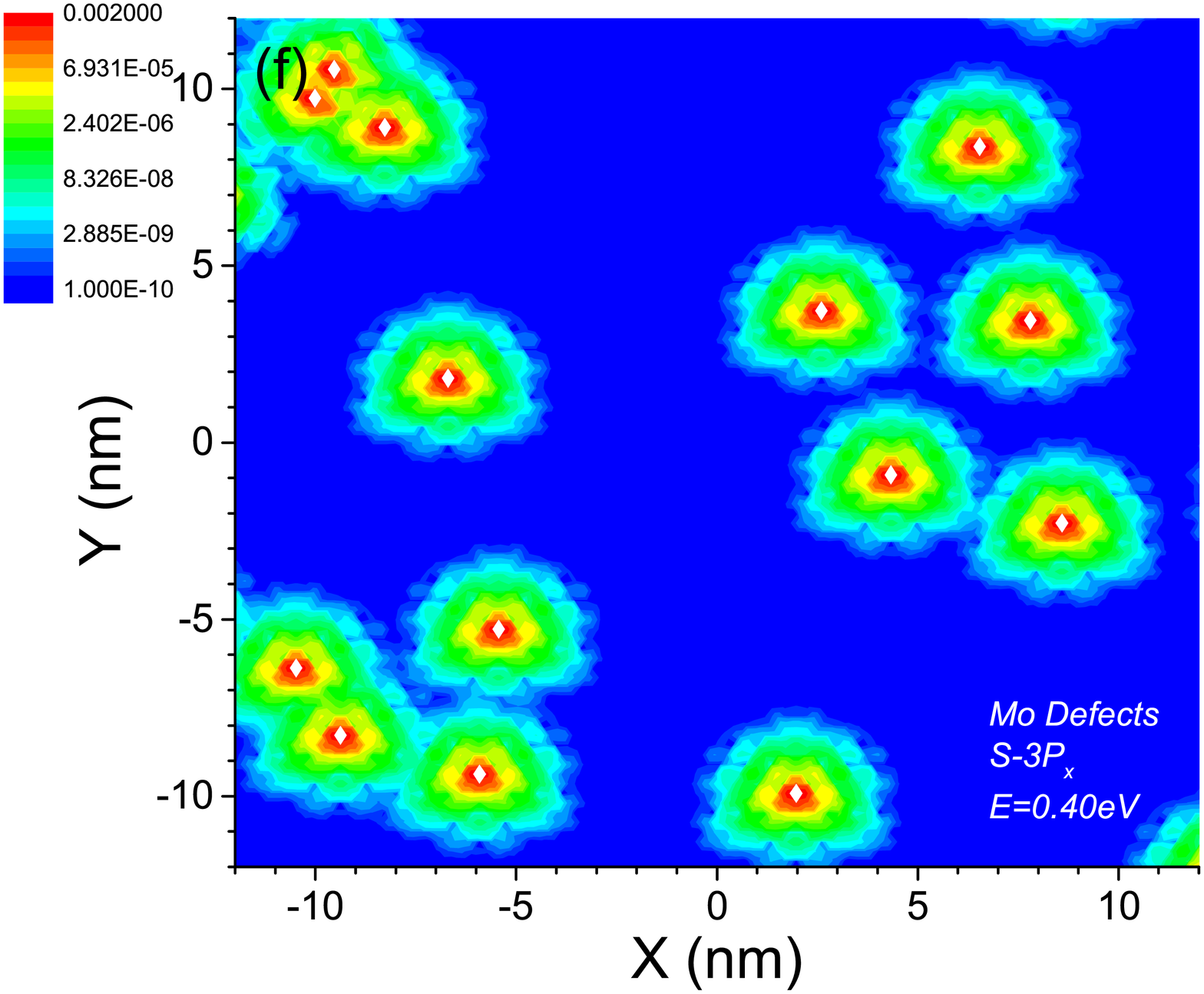}
}
\end{center}
\caption{LDOS of for the corresponding dominant orbitals in the impurity states of MoS$_2$ with single point defects. The white points indicate the position of the defects. The concentration of the defects is $0.1\%$ for all panels. Results are obtained from the average of quasi-eigenstates for one hundred samples with fixed distribution of point defects but different initial states. Please notice the different color scale in each panel.
}
\label{Fig:EigenState}
\end{figure*}

{\it Method---} Transition metal dichalcogenides as MoS$_2$ and WS$_2$ are composed, in its bulk configuration, of two-dimensional S-$M$-S layers ($M=$Mo,W) stacked on top of each other, coupled by weak van der Waals forces. The transition metal atoms $M$ are ordered in a triangular lattice, each of them bonded to six S atoms located in the top and bottom layers, forming a sandwiched material. Similarly as in graphene, the weak interlayer coupling makes possible to exfoliate this material down to a single-layer \cite{NG05}. The electronic band structure of MoS$_2$ changes from an indirect band gap for multilayer samples, to a direct gap semiconductor for single-layers \cite{SW10,CG13}. We consider a 6-bands tight-binding model which contains the proper orbital combination that contributes to the valence and conduction bands of $M$S$_2$: 3 $d$-orbitals of the transition metal ($d_{xy}$, $d_{x^2-y^2}$ and $d_{3z^2-r^2}$) as well as the symmetric (antisymmetric) combination of the $p_x,p_y$ ($p_z$) orbitals of the top and bottom chalcogen atoms \cite{CG13,CS13}. The base vector can be written as
\begin{equation}
\phi_i^\dagger
=
(
d_{i,3z^2-r^2}^\dagger,
d_{i,x^2-y^2}^\dagger,
d_{i,xy}^\dagger,
p_{i,x,A}^\dagger,
p_{i,y,A}^\dagger,
p_{i,z,S}^\dagger
),
\label{basis}
\end{equation}
where $p^{\dagger}_{i,o,S}=(p^{\dagger}_{i,o,t}+p^{\dagger}_{i,o,b})/\sqrt{2}$, $p^{\dagger}_{i,o,A}=(p^{\dagger}_{i,o,t}-p^{\dagger}_{i,o,b})/\sqrt{2}$, $o=x,y,z$ and the subscripts $t$ and $b$ refer to the top and bottom S layers, respectively.

We also consider the intra-atomic spin-orbit coupling $\sum_a\lambda_a {\bf \hat{L}}_a\cdot{\bf \hat{S}}_a$, where $a=M,$ S accounts for both, the transition metal $M$ as well as the chalcogen atom S, $\lambda_a$ is the corresponding intra-atomic SO interaction, ${\bf \hat{L}}$ is the atomic angular momentum operator, and ${\bf \hat{S}}=\hbar{\boldsymbol {\hat{\sigma}}}$  is the spin operator. 
The optical and electronic properties, such as density of states (DOS), quasi-eigenstates, optical and DC conductivities, are obtained numerically by using the TBPM \cite{YRK10,WK10,YRRK11,Yuan2012} (more details can be found in the Supplementary Material).

{\it Results and discussion---} The effect of point defects in the DOS of MoS$_2$ and WS$_2$ are shown in Fig. \ref{Fig:DOS&Optical}(a)-(c). The defect concentration $(0.2\%)$ for single point defects in Fig. \ref{Fig:DOS&Optical}(a) and (b) is chosen to be of the same order as the intrinsic vacancies observed in recent experiments \cite{QiuH2013}. For clean samples (black lines), the DOS has a gap $\Delta$ which corresponds to the well known direct gap of single layer samples at the $K$ points of the Brillouin zone (BZ). Defects in the samples lead to the appearance of a series of peaks in the gapped region of the DOS, which are associated to the creation of midgap states localized around the defects, whose energy and strength depends on the specific missing atoms, their concentration as well as the specific arrangement of the point defects as individual missing atoms [Fig. \ref{Fig:DOS&Optical}(a) and (b)] or in clusters of point defects with variable radius [Fig. \ref{Fig:DOS&Optical}(c)]. For the same concentration of defects, isolated point defects modify more strongly the DOS than clusters of defects. This is the reason why we show results for 0.2\% of single defects, and 1\% of cluster of defects with a radius $R=a$ \footnote{See Supplementary Material for a detailed description of the arrangement of the defects in our calculation.}. The impurity states have also an important effect on the optical conductivity [Fig. \ref{Fig:DOS&Optical}(d)-(e)]. First, let us consider the case of undoped and clean MoS$_2$ and WS$_2$. Since single layers of those TMD are direct gap semiconductors, the only optical transitions allowed at low energies are two set of inter-band transitions with $\omega\ge \Delta$ from the edges of the SOC split valence bands to the conduction band at the $K$ and $K'$ points of the BZ \cite{LC12}. Those transitions lead to the A and B absorption peaks observed in photoluminescence experiments \cite{MH10}, and the SOC splitting of the valence band manifests itself in the optical conductivity through the step like feature of $\sigma(\omega)$ that can be seen in the black lines (for pristine MoS$_2$ and WS$_2$) of Fig. \ref{Fig:DOS&Optical}(d)-(f). This feature is especially visible for WS$_2$ due to the strong SOC associated to the heavy W atom \cite{RG14}, which lead to a plateau like feature for $\sigma(\omega)$ of $\sim400$ meV, in agreement with the energy separation between the spin polarized valence bands. The existence of defects in the sample lead to flat midgap bands which activate new optical transitions with $\omega<\Delta$ in the optical spectrum. Most importantly, these new optical transitions lead to a background contribution which appears in the optical conductivity at low energies, as it can be seen in Fig. \ref{Fig:DOS&Optical}(d)-(f) for different concentration of defects, suggesting that resonant impurities, like the defects studied in this work, could have a relevant contribution to the optical spectrum of TMDs \cite{MH10}. Furthermore, it is interesting to note that this background contribution due to disorder resembles that observed in the the optical conductivity of highly doped graphene \cite{LiZQ2008}.

Our results for the DC conductivities are shown in Fig. \ref{Fig:DC}, which demonstrate a significant asymmetry between electrons and holes, in reasonable agreement with experiments \cite{BF13} (note  that the observations are done in multilayered samples). The fact that Mo and W defects lead to localized states well inside the gap, as illustrated by the densities of states also shown in Fig. \ref{Fig:DC}, combined with the fact that the bands at the $K$ and $K'$ points of the BZ can be approximated by an effective gapped Dirac equation \cite{XY12,LS13}, suggests that these defects can act as resonant scatterers \cite{SPG07}, which give rise to a mobility almost independent of the carrier density (see below and the T-matrix analysis in the Supplementary Material). 

The localization of these midgap states are clearly seen by their local
density of states (LDOS) plotted in Fig. \ref{Fig:EigenState}. The amplitude
of each orbital in real space is obtained from the average of
quasiegenstates with different initial states. The profiles of the localized
states show either hexagonal symmetry or mirror symmetry, depending on the
type of the orbital and the energy of the midgap states. For S defects,
the impurity state at the energy $-0.86$~eV is localized mainly ($\sim 65.6\%$) on $%
d_{3z^{2}-r^{2}}$ orbitials of neighboring Mo atoms, with small
amount ($\sim 9.6\%$) on $p_{z,S}$ orbitals of neighboring S atoms. For Mo
defects, on the other hand, there are two midgap states, one centered at $0.18$~eV, with
mainly localized $p_{x(y),A}\ (\sim 23.4\%)$ and $d_{3z^{2}-r^{2}}$ $(\sim 21.2\%)$ orbitals, and
another centered at $0.4$~eV, with mainly localized $d_{3z^{2}-r^{2}}$ $(\sim 46.4\%)$
and $p_{x(y),A}\ (\sim 21.6\%)$ orbitals. The detailed DOS of each orbital as a
function of energy is shown in the Supplementary Material.

\begin{figure}[t]
\begin{center}
\mbox{
\includegraphics[width=7cm]{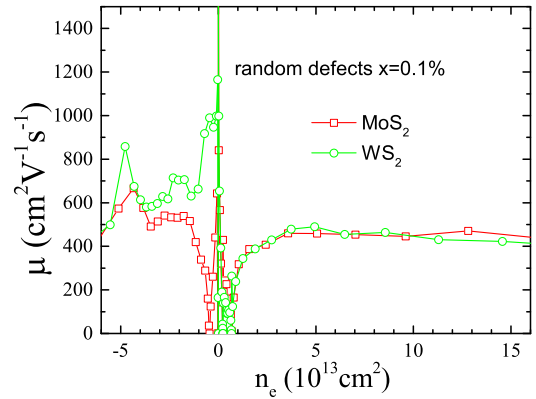}
}
\end{center}
\caption{Mobility for MoS$_2$ and WS$_2$ for the same concentration (0.1\%) of point defects. $n$-doping corresponds to Mo(W) point defects and $p$-doping corresponds to S point defects. }
\label{Fig:Mobility}
\end{figure}

Results for the carrier mobility, defined as $\mu
\left( E\right) =\sigma \left( E\right) /en_{e}\left( E\right) $, where
the charge density $n_{e}\left( E\right) $ is obtained from the integral of
density of states via $n_{e}\left( E\right) =\int_{0}^{E}\rho \left(
\varepsilon \right) d\varepsilon $ are shown in Fig. \ref{Fig:Mobility}. We notice that $n$-doping corresponds to Mo(W) point defects, whereas $p$-doping corresponds to S point defects. We observe that for $n$-doped samples, MoS$_2$ and WS$_2$ show similar mobilities, whereas for $p$-doped samples, the mobility of WS$_2$ is larger than for MoS$_2$. Our results show that in general, the mobilities of TMDs are low, but they are larger for holes than for electrons, in agreement with previous experimental results \cite{BF13,ZA14}.
The results for the mobility suggest that it is independent of carrier concentration, except at the edge of the valence band. As we have discussed before, this is consistent with the expected features of resonant scatterers. A more detailed analysis of resonant scatterers in gapped Dirac systems is required in order to make this statement more quantitative, however. Note, finally, that our analysis leaves out the effect of the missing charge at the defect, which can lead to a long range potential, and to intravalley scattering \cite{P10}.

{\it Conclusions---} We have studied the effect of point defects in the DOS, optical and DC conductivity of single layers of TMDs like MoS$_2$ and WS$_2$. The existence of point defects in the sample creates flat midgap bands which activate new optical transitions in the optical spectrum, leading to a background contribution which appears in the optical conductivity at low energies, in agreement with photoconductivity measurements. Our results show a significant asymmetry between electrons and holes. The DC conductivities and mobilities are larger for holes, in agreement with experiments, and we find higher mobilities for $p$-doped WS$_2$ than for MoS$_2$. Mo and W defects induce localized states well inside the gap, suggesting a behavior similar to that of resonant scatterers in graphene. 

{\it Acknowledgments---}
We thank the European Union Seventh Framework Programme under grant agreement n604391 Graphene Flagship. The support by the Stichting Fundamenteel Onderzoek der Materie (FOM) and
the Netherlands National Computing Facilities foundation (NCF) are
acknowledged. S.Y. and M.I.K. thank financial support from the
European Research Council Advanced Grant program
(contract 338957). RR acknowledges financial support from the Juan de la Cierva Program (MEC, Spain). R.R. and F.G. thank financial support from MINECO, Spain, through Grant No. FIS2011-23713, and the European Research Council Advanced Grant program (contract 290846).

\bibliographystyle{apsrev}
\bibliography{Biblio_MoS2_Disorder}

\newpage

{\bf Supplementary Material}

\begin{figure}[h]
\begin{center}
\mbox{
\includegraphics[width=4.2cm]{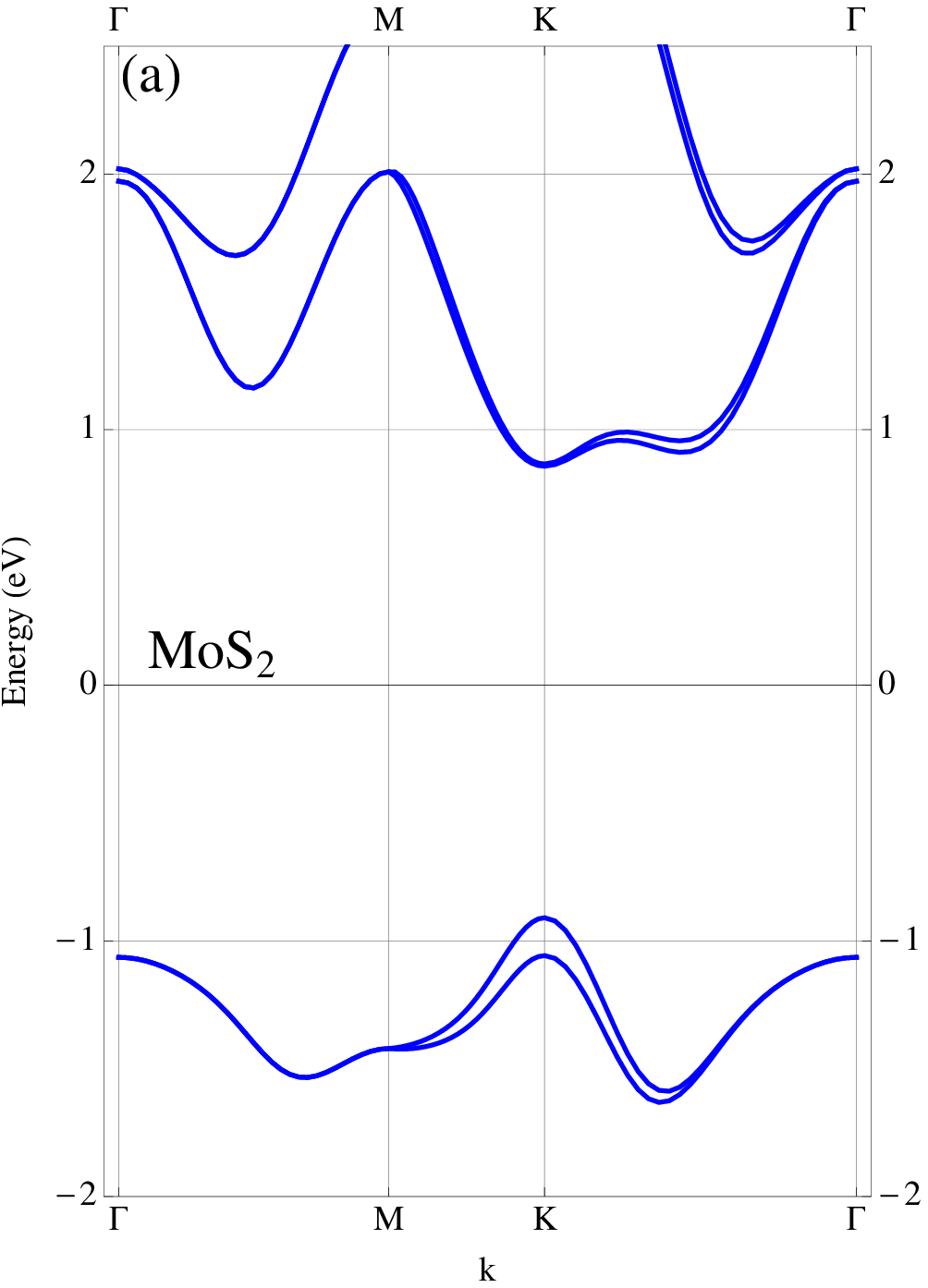}
\includegraphics[width=4.2cm]{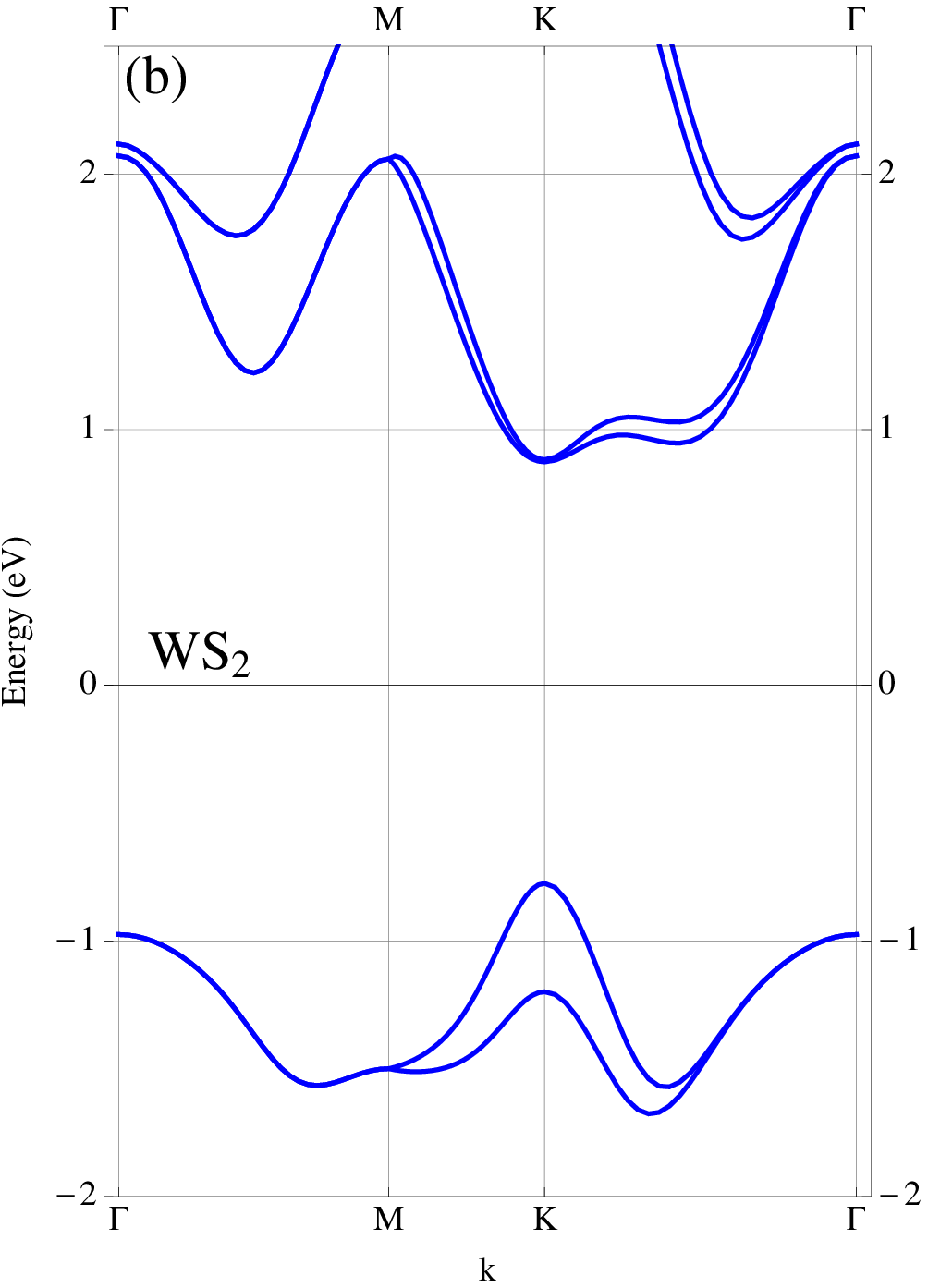}
}
\end{center}
\caption{Band structure of MoS$_2$ (a) and WS$_2$ (b) obtained by the TB model used in the text. The Slater-Koster TB parameters are those given in Ref. \cite{RG14} }
\label{Fig:Bands}
\end{figure}

{\it Tight-binding band structure and distribution of defects---} In Fig. \ref{Fig:Bands} we show the band structure of MoS$_2$ and WS$_2$ obtained from the TB model used in the calculations. The TB parameters are given in Ref. \cite{RG14}.

\begin{figure}[h]
\begin{center}
\includegraphics[width=8.5cm]{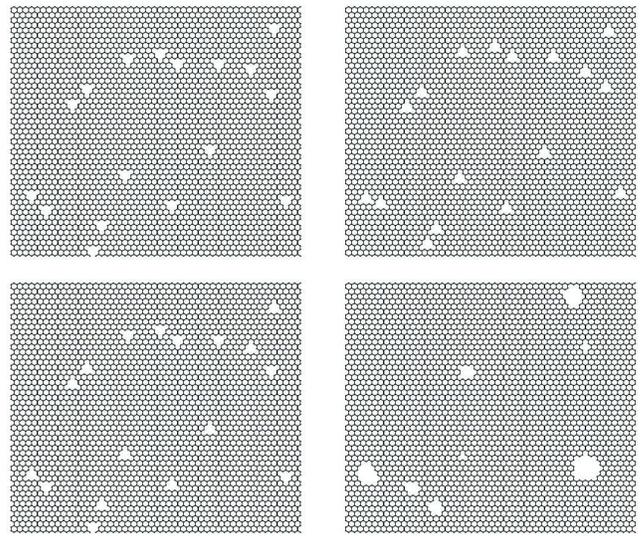}
%\mbox{
%\includegraphics[width=4.2cm]{xcluster_Mo.eps}
%\includegraphics[width=4.2cm]{xcluster_S.eps}
%}
%\mbox{
%\includegraphics[width=4.2cm]{xcluster_Mo_S.eps}
%\includegraphics[width=4.2cm]{xcluster_random.eps}
%}
\end{center}
\caption{Sketch of a MoS$_2$ or WS$_2$ sheet with defect-clusters. Top left: $R=a$ with cluster centers on Mo (MoS6 defects); top right: $R=a$ with cluster centers on S (Mo3S2 defects); bottom left: $R=a$ with cluster centers on Mo and S (MoS6 and Mo3S2 defects); bottom right: random defect-clusters ($R<3.5a$), as described in the text. For illustrative purposes, the size of the sample shown in this sketch is $50 \times 100$ (considerably smaller than the sizes used in our simulations), and the concentration of defects is approximately equal to 1\%. }
\label{Fig:Samples}
\end{figure}

In Fig. \ref{Fig:Samples} we show an sketch of the distribution of defects considered in the main text.

\bigskip
{\it Tight-binding Propagation Method---} Our method is based on the numerical solution
of time-dependent Sch\"{o}dinger equation (TDSE) in the TB model. The initial state $\left\vert \varphi \right\rangle $ is considered as a random
superposition of all orbitals over the whole space which covers all the
energy eigenstates \cite{HR00,YRK10}%
\begin{equation}
\left\vert \varphi \right\rangle =\sum_{i,o,\sigma}a_{i,o,\sigma}\left\vert
i,o,\sigma\right\rangle ,  \label{Eq:phi0}
\end{equation}%
where $a_{i,o,\sigma}$ are random complex numbers normalized as $\sum_{i,o,\sigma}\left%
\vert a_{i,o,\sigma}\right\vert ^{2}=1$, and $\left\vert i,o,\sigma\right\rangle $
represents the $o$ orbital with spin $\sigma$ at site $i$. The density of states can be obtained by
the Fourier transformation of the overlap between the time-evolved state $\left\vert
\varphi (t)\right\rangle \equiv e^{-i\mathcal{H}t}\left\vert \varphi
\right\rangle $ and the initial state $\left\vert \varphi \right\rangle $ as
\cite{HR00,YRK10}
\begin{equation}
\rho \left( \varepsilon \right) =\frac{1}{2\pi }\int_{-\infty }^{\infty
}e^{i\varepsilon t}\left\langle \varphi |\varphi (t)\right\rangle dt.
\label{Eq:DOS}
\end{equation}
Here we use units such that $\hbar =1$. The time-evolution operator $e^{-i%
\mathcal{H}t}$ is calcualted numerically by using Chebyshev polynomial
algorithm, extremly efficient for a TB Hamitlonian $\mathcal{H}$ which is a
sparse matrix. Within the TBPM, the optical conductivity (omitting the Drude contribution at $\omega =0$%
) is calculated by using the Kubo formula \cite{Ishihara1971,YRK10}
\begin{eqnarray}
\sigma _{\alpha \beta }\left( \omega \right)  &=&\lim_{\epsilon \rightarrow
0^{+}}\frac{e^{-\tilde{\beta} \omega }-1}{\omega \Omega }\int_{0}^{\infty
}e^{-\epsilon t}\sin \omega t  \notag  \label{gabw2} \\
&&\times 2~\text{Im}\left\langle \varphi |f\left( \mathcal{H}\right)
J_{\alpha }\left( t\right) \left[ 1-f\left( \mathcal{H}\right) \right]
J_{\beta }|\varphi \right\rangle dt,  \notag \\
&&
\label{Eq:OptCond}
\end{eqnarray}%
where $\tilde{\beta} =1/k_{B}T$ is the inverse temperature, $\Omega $ is the sample
area, $f\left( \mathcal{H}\right) =1/\left[ e^{\tilde{\beta} \left( \mathcal{H}-\mu
\right) }+1\right] $ is the Fermi-Dirac distribution operator, and the
time-dependent current operator in the $\alpha $ ($=x$ or $y$) direction is $%
J_{\alpha }\left( t\right) =e^{i\mathcal{H}t}J_{\alpha }e^{-i\mathcal{H}t}$.

The DC conductivity at zero temperature is calculated by using the Kubo
formula at $\omega \rightarrow 0$ \cite{Ishihara1971,YRK10}
\begin{eqnarray}
\mathbf{\sigma }_{\alpha \alpha } &=&\frac{\rho \left( \varepsilon \right) }{%
\Omega }\int_{0}^{\infty }dt~\text{Re}\left[ e^{-i\epsilon t}\left\langle
\varphi \right\vert J_{\alpha }e^{i\mathcal{H}t}J_{\alpha }\left\vert
\varepsilon \right\rangle \right] ,  \notag \\
&&
\end{eqnarray}%
where $\left\vert \varepsilon \right\rangle $ is the {\it normalized}
quasi-eigensate
\footnote{
Following Ref. \cite{YRK10}, the {\it normalized} quasi-eigenstate $\left\vert \varepsilon \right\rangle =\left\vert \Phi
\left( \varepsilon \right) \right\rangle /\left\vert \left\langle \varphi
|\Phi \left( \varepsilon \right) \right\rangle \right\vert $ is defined from
the quasi-eigenstate $\left\vert \Phi \left( E\right) \right\rangle $, which
is a superposition of the degenerate eigenstates with the same eigenenergy $E
$, obtained as the Fourier transform of $\left\vert \varphi (t)\right\rangle $, i.e.
$
\left\vert \Phi \left( E\right) \right\rangle =\frac{1}{2\pi }\int_{-\infty
}^{\infty }dte^{iEt}\left\vert \varphi \left( t\right) \right\rangle .
$
The quasi-eigenstate is not exactly an energy eigenstate, unless the
corresponding eigenstate is not degenerate at energy $E$. The average of the real space distribution of the amplitude for different realization of intial states can be used to examine the
localization of the modes \cite{YRK10,YRRK11,Yuan2012}.
}.
The accuracy of TBPM is mainly determined by the time interval and total time steps used in the Fourier transformation. The main limitation of the numerical calculations is the size of the physical memory that can be used to store the quasi-eigenstates $\left\vert \Phi \left(E\right) \right\rangle $ in the calculation of DC conductivity. In the present work, we have fixed the temperature to $T=300$~K for the optical conductivity and to $T=0$ for the DC conductivity. We study systems containing $2400\times 2400$ atoms, with periodic boundary conditions.

\begin{figure}[t]
\begin{center}
\mbox{
\includegraphics[width=8cm]{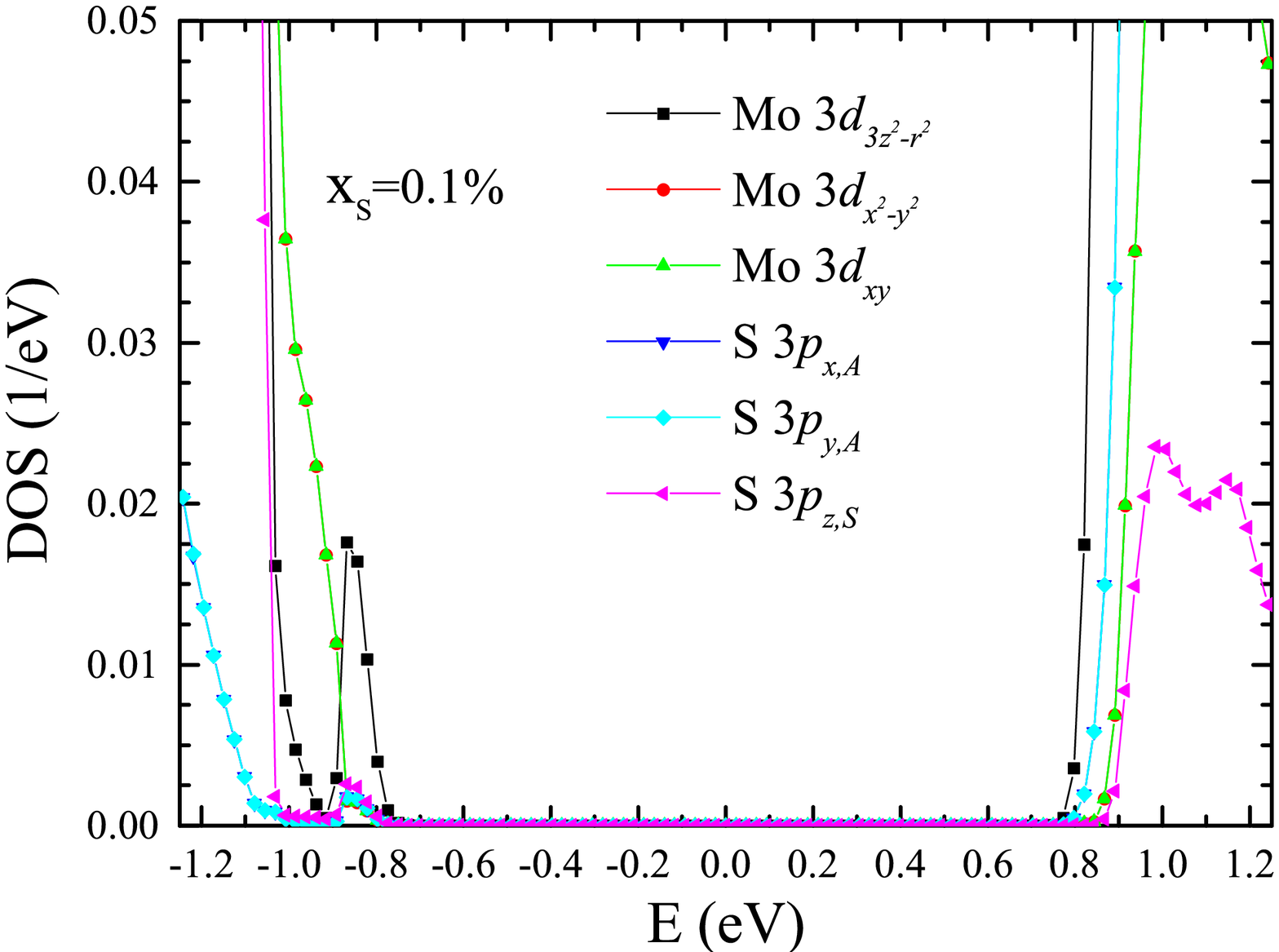}
}
\mbox{
\includegraphics[width=8cm]{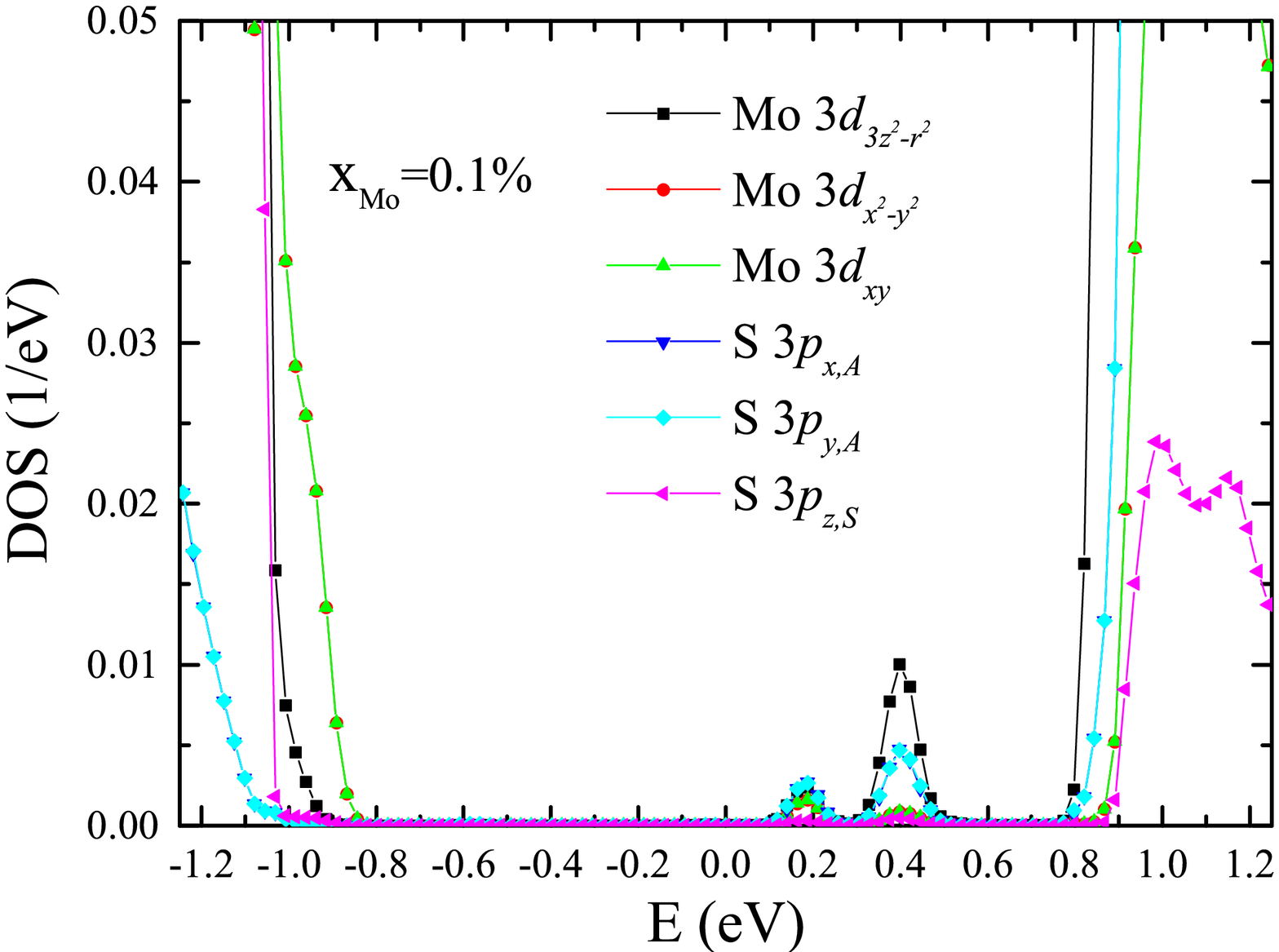}
}
\end{center}
\caption{Orbital resolved DOS for MoS$_2$ with 0.1\% of S defects (top panel) or Mo defects (bottom panel). See also Fig. 3 of the main text.}
\label{Fig:OrbitalDOS}
\end{figure}

\bigskip
{\it Orbital DOS---} Here we complement the information for the localized states around the impurities shown in Fig. 3 of the main text, and we show in Fig. \ref{Fig:OrbitalDOS} the contribution of each orbital to the DOS of MoS$_2$ with 0.1\% of S or Mo defects.

\begin{figure}[h]
\begin{center}
\mbox{
\includegraphics[width=4.2cm]{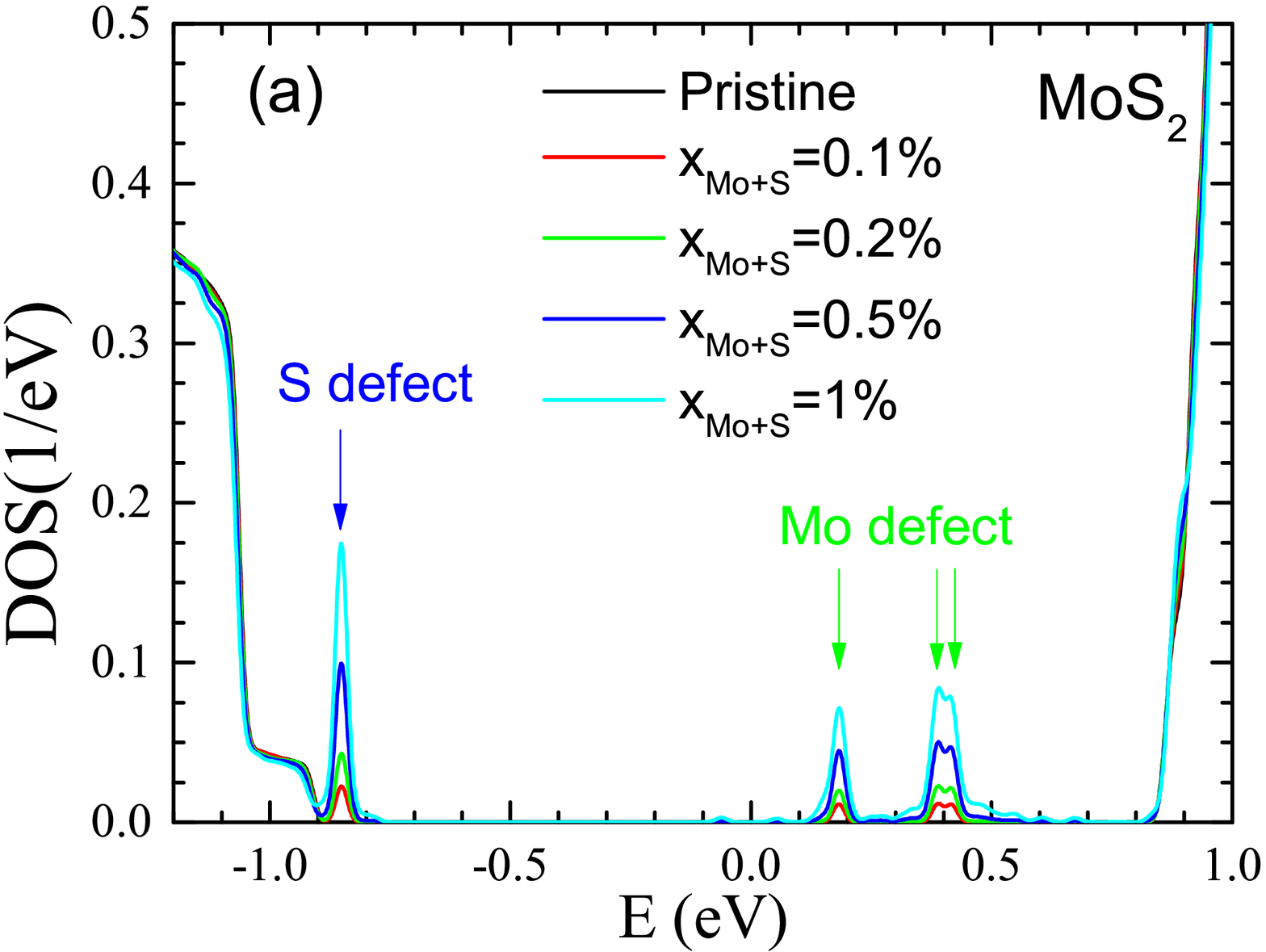}
\includegraphics[width=4.2cm]{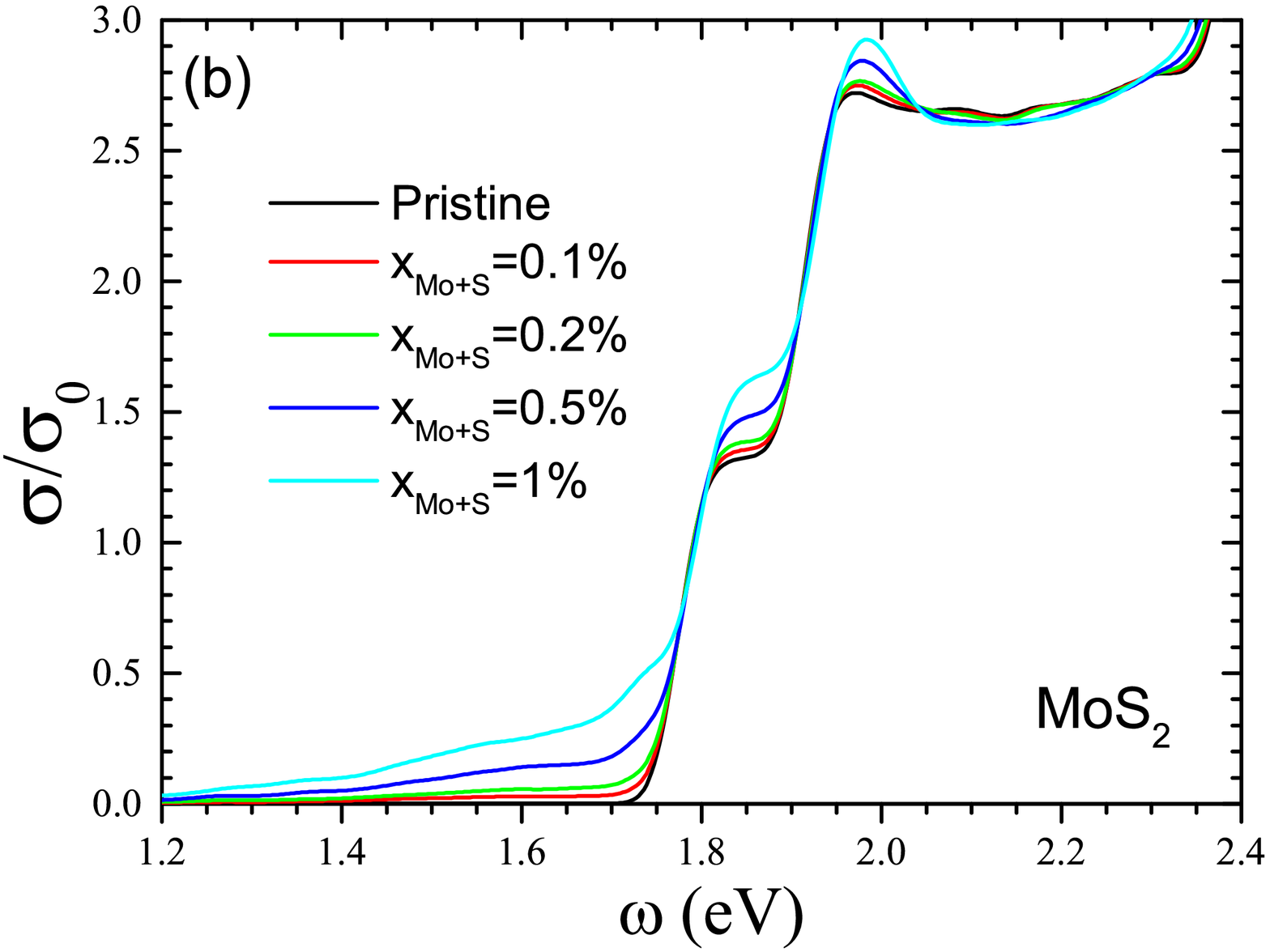}
}
\mbox{
\includegraphics[width=4.2cm]{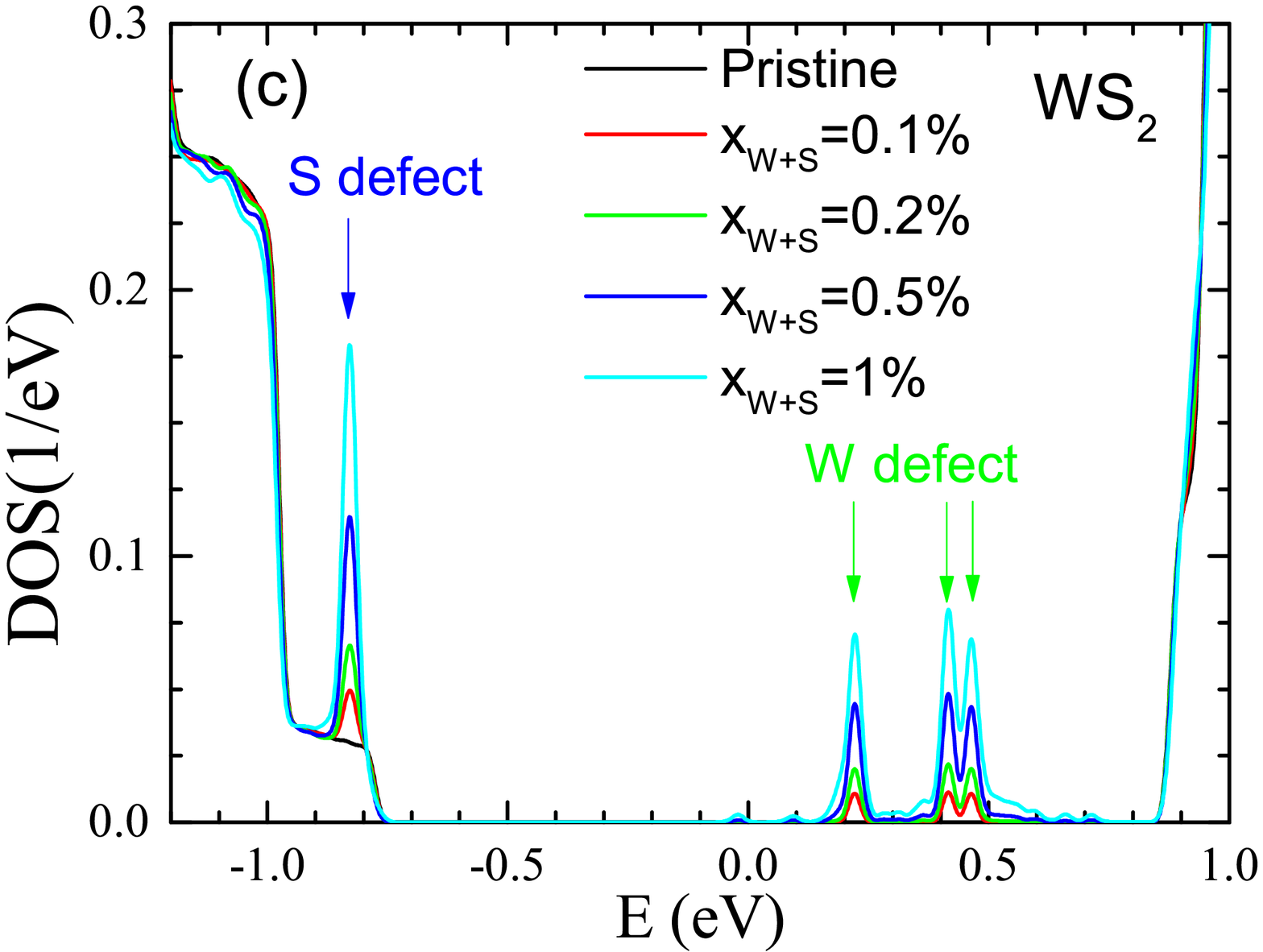}
\includegraphics[width=4.2cm]{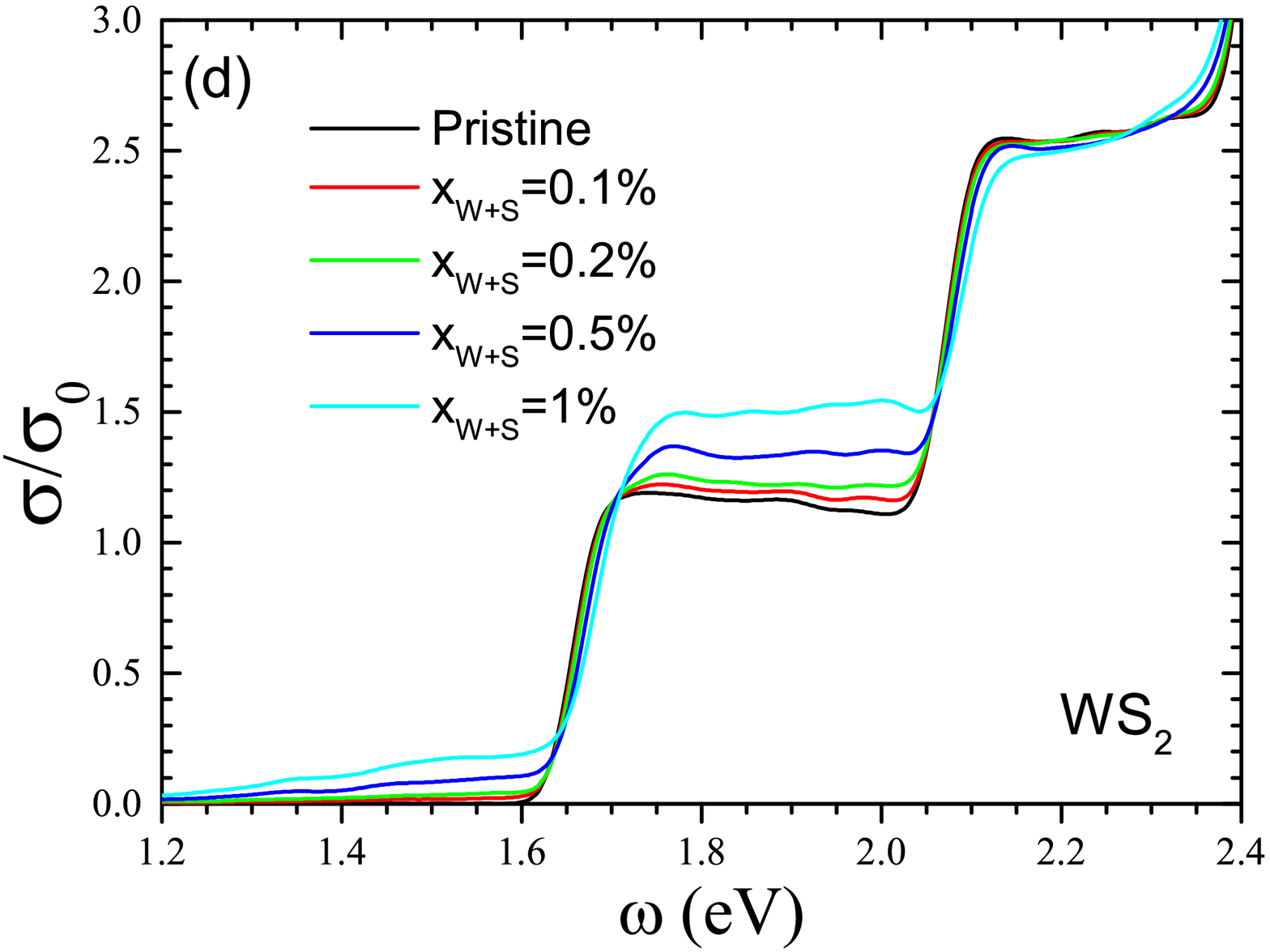}
}
\end{center}
\caption{Comparison of DOS and optical conductivity of MoS$_2$ (a) and WS$_2$ (b) for different concentrations of defects.}
\label{Fig:DOS&OpticalComparison}
\end{figure}

\bigskip
{\it Dependence of the conductivity on the concentration of defects---} In Fig. \ref{Fig:DOS&OpticalComparison} we show results similar to Fig. 1 of the main text, but comparing different concentrations of defects. The height of the peaks in the DOS in the middle of the gap, associated to localized states around the impurities, increases with disorder, resulting in a larger background contribution to the optical conductivity for energies lower than the gap. 

\begin{figure}[h]
\begin{center}
\mbox{
\includegraphics[width=6cm]{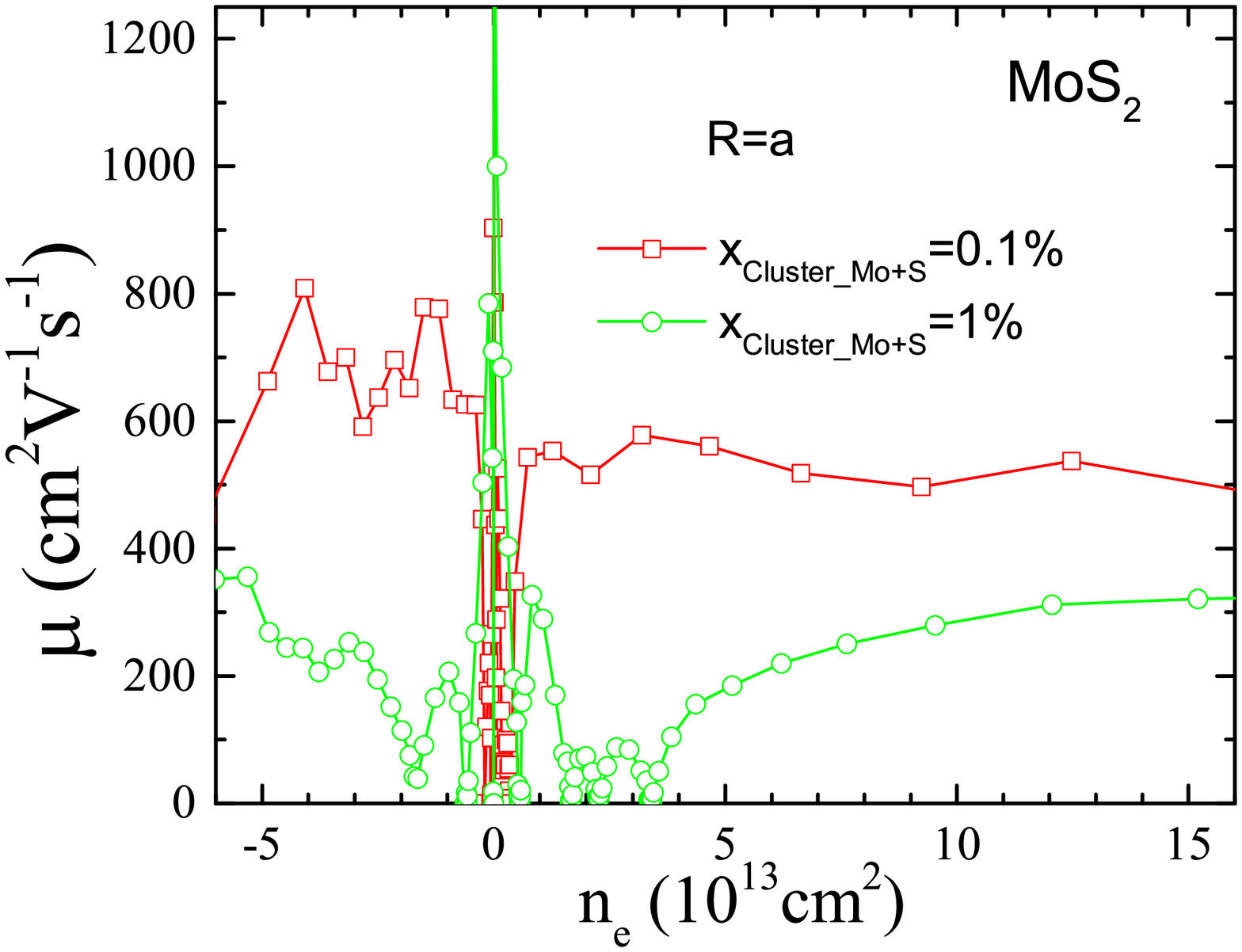}
}
\mbox{
\includegraphics[width=6cm]{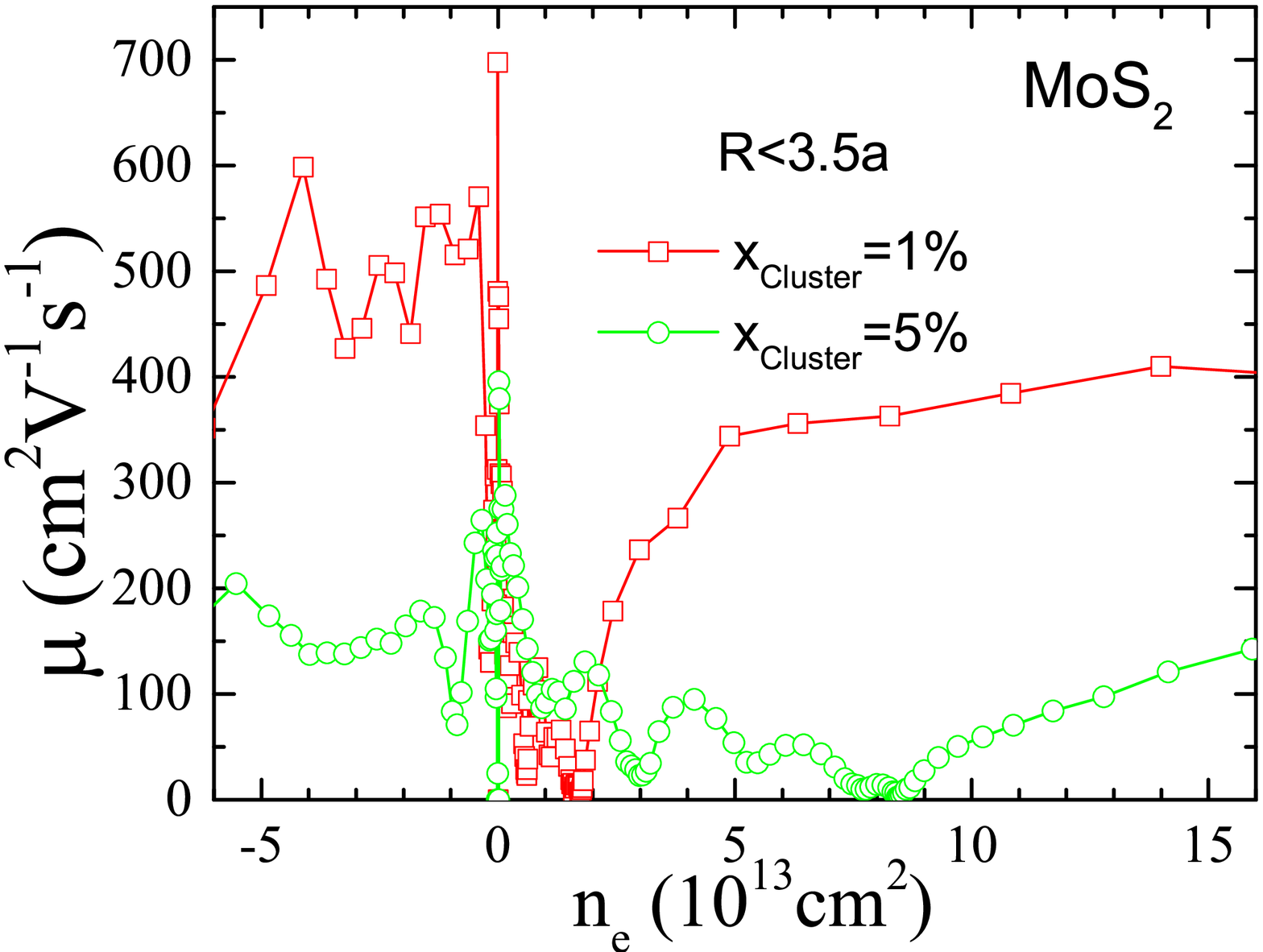}
}
\end{center}
\caption{Comparison of mobility of MoS$_2$ for different concentrations of cluster defects: (a) cluster centers located on Mo or S with $R=a$, and (b) random clusters with $R<3.5a$.}
\label{Fig:mobility_cluster}
\end{figure}

The mobility of the samples decreases with the concentration of defects, as it is shown in Fig. \ref{Fig:mobility_cluster} for samples with clusters of defects, as stated in the figure.

%\bigskip
{\it Low energy model for the DC conductivity---}
The numerical results shown in Fig. 2 for the DC conductivity can be complemented with a low energy approximation in which we can calculate the conductivity using the $T$ matrix \cite{WKL09}, which accounts for the scattering of electrons by resonant impurities
\begin{equation}
T(E)=\frac{V^2}{E-\epsilon_d-V^2g_0(E)}
\end{equation}
where $V$ is the potential accounting for the impurity, and $g_0(E)$ is the local unperturbed Green's function, which for a semiconductor with electron-hole asymmetry can be obtained from a density of states of the form $N_0(E)=D_c\Theta(E-\Delta/2)+D_v\Theta(-\Delta/2-E)$, from which we obtain
\begin{equation}
g_0(E)=D_c\log\left|\frac{E-\Delta/2}{E-W_c}\right|+D_v\log\left|\frac{E+W_v}{E+\Delta/2}\right|
\end{equation}
where $W_{c(v)}$ accounts for the width of the conduction (valence) bands. The case of interest here, which is defects, can be considered by the limit $V\rightarrow \infty$ which leads to $T\rightarrow-1/g_0(E)$. From this the conductivity can be calculated from $\sigma=(2e^2/h)E_F\tau$ where $\tau^{-1}=(2\pi/\hbar)n_i|T(E)|^2N_0(E_F)$ is the scattering rate in terms of the concentration of impurities $n_i$.

%\iffalse

%\newcommand{\npb}{Nucl. Phys.}\newcommand{\adv}{Adv.
%  Phys.}\newcommand{\epl}{Europhys. Lett.}

%\fi

%\bibliographystyle{apsrev}
%\bibliography{Biblio_MoS2_Disorder}

\end{document}